\documentclass[preprint,prd,
amsmath,amssymb,amsthm,nofootinbib,superscriptaddress,longbibliography]{revtex4-1}

\usepackage{graphicx}   
\usepackage[
colorlinks=true,        
citecolor=blue,         
linkcolor=blue,         
urlcolor=blue ,          
]{hyperref}  
\usepackage{tabulary}
\usepackage{color}      
\usepackage{orcidlink}
\usepackage{multirow}
\usepackage{floatrow}
\usepackage{float} 

\newcommand{\nc}{\newcommand*} 
\newcommand{\be}{\begin{equation}}
\newcommand{\ee}{\end{equation}}
\newcommand{\bea}{\setlength\arraycolsep{2pt} \begin{eqnarray}}
\newcommand{\eea}{\end{eqnarray}}

\nc{\al}{\alpha}
\nc{\s}{\sigma}
\nc{\dt}{\delta}
\nc{\Dt}{\Delta}
\nc{\Ld}{\Lambda}
\nc{\p}{\partial}
\nc{\om}{\omega}
\nc{\Om}{\Omega}
\nc{\rd}{\mathrm{d}}
\nc{\Od}[1]{\mathcal{O}(#1)} 
\nc{\kp}{\kappa}

\def\[{\left[}
\def\]{\right]}
\def\e{\begin{equation}}
\def\q{\end{equation}}
\def\m{\begin{eqnarray}}
\def\n{\end{eqnarray}}

\nc{\Eq}[1]{Eq.~\eqref{#1}}     
\nc{\Fig}[1]{Fig.~\ref{#1}}     
\nc{\Table}[1]{Table~\ref{#1}}  
\nc{\Sec}[1]{Sec.~\ref{#1}}     

\nc{\Msun}{M_\odot}             
\nc{\fpbhn}{f_{\mathrm{pbh0}}}    
\nc{\mR}{\mathcal{R}} 
\nc{\seq}{\sigma_{\mathrm{eq}}}
\nc{\ogw}{\Omega_{\mathrm{GW}}}
\nc{\gpcyr}{\mathrm{Gpc}^{-3}\,\mathrm{yr}^{-1}}
\nc{\lvc}{LIGO/Virgo} 
\nc{\SNR}{\mathrm{SNR}} 
\nc{\mmin}{{m_{\mathrm{min}}}}
\nc{\mmax}{{m_{\mathrm{max}}}}
\nc{\Mmin}{{M_{\mathrm{min}}}}
\nc{\fmin}{{f_{\mathrm{min}}}}
\nc{\VT}{\mathrm{VT}}
\nc{\rhoGW}{\rho_{\mathrm{GW}}}
\nc{\vth}{\vec{\theta}}
\nc{\vd}{\vec{d}}
\nc{\vla}{\vec{\lambda}}
\nc{\Nobs}{N_{\mathrm{obs}}}
\nc{\av}[1]{\langle #1 \rangle} 
\nc{\km}{\mathrm{km}}
\nc{\Mpc}{\mathrm{Mpc}}
\nc{\Tobs}{T_{\mathrm{obs}}}
\nc{\Ntemp}{N_{\mathrm{temp}}}
\nc{\ie}{\textit{i.e.}}
\nc{\eg}{\textit{e.g.~}}
\nc{\app}{\approx}
\nc{\hf}{\frac{1}{2}}

\nc{\mpbh}{m_{\rm{pbh}}}
\nc{\cR}{\mathcal{R}}
\nc{\mU}{{\mathcal{U}}}
\nc{\Mc}{{M_\mathrm{c}}}
\nc{\Mf}{{M_\mathrm{f}}}
\nc{\red}[1]{\textcolor{red}{#1}}
\nc{\yellow}[1]{\textcolor{yellow}{#1}}
\nc{\green}[1]{\textcolor{green}{#1}}
\nc{\blue}[1]{\textcolor{blue}{#1}}

\begin{document}
\title{Probability density function for  dispersion measure of  fast radio burst  from extragalactic medium} 

\author{Yuchen Zhang\orcidlink{0009-0009-3644-2082}
}
\affiliation{Department of Physics, Institute of Interdisciplinary Studies and Synergetic Innovation Center for Quantum Effects and Applications, Hunan Normal University, Changsha, Hunan 410081, China}

\author{Yang Liu\orcidlink{0000-0003-2721-2559}
}
\email{Co-first author.}
\affiliation{Department of Physics, Institute of Interdisciplinary Studies and Synergetic Innovation Center for Quantum Effects and Applications, Hunan Normal University, Changsha, Hunan 410081, China}

\author{Hongwei Yu\orcidlink{0000-0002-3303-9724}
}
\email{hwyu@hunnu.edu.cn}
\affiliation{Department of Physics, Institute of Interdisciplinary Studies and Synergetic Innovation Center for Quantum Effects and Applications, Hunan Normal University, Changsha, Hunan 410081, China}
\affiliation{Hunan Research Center of the Basic Discipline for Quantum Effects and Quantum Technologies, Hunan Normal University, Changsha, Hunan 410081, China}

\author{Puxun Wu\orcidlink{0000-0002-9188-7393}
}
\email{pxwu@hunnu.edu.cn}
\affiliation{Department of Physics, Institute of Interdisciplinary Studies and Synergetic Innovation Center for Quantum Effects and Applications, Hunan Normal University, Changsha, Hunan 410081, China}
\affiliation{Hunan Research Center of the Basic Discipline for Quantum Effects and Quantum Technologies, Hunan Normal University, Changsha, Hunan 410081, China}

\begin{abstract}
  
  Fast Radio Bursts (FRBs) have emerged as powerful probes in cosmology.  An optimized method  was recently proposed to  extract the cosmic baryon density from localized FRBs by maximizing the joint likelihood function of the extragalactic dispersion measure ($\mathrm{DM}_{\mathrm{ext}}$).  
  In this paper, we  identify a crucial factor  that was omitted in 
   the probability density function (PDF) for $\mathrm{DM}_{\mathrm{ext}}$    in that method. Using simulated FRB data, we demonstrate that neglecting this factor leads to a systematic bias in the inferred cosmic baryon density, with deviations exceeding the $1\sigma$ confidence level. This highlights the necessity of including the missing factor for reliable cosmological applications of FRBs.  Furthermore,  applying our corrected PDF to a sample of 88 real localized FRBs, we find that the baryon density inferred with the original PDF is inconsistent with the Planck 2018 CMB results, whereas our corrected PDF yields excellent agreement.
 \end{abstract}

\maketitle

\section{Introduction}

Fast Radio Bursts (FRBs) are intense bursts of radio waves lasting just milliseconds, yet releasing energy comparable to what the Sun emits over several days. They were first discovered in archival data in 2007~\cite{lorimer2007bright}, and now are observed across the entire sky, with thousands of events estimated to occur daily. Although their exact origin remains unknown, it is widely accepted that many FRBs have cosmological origins since their observed dispersion measures (DM), a quantity representing the integrated free-electron density along the line of sight, are typically much higher than expected from contributions solely within the Milky Way~\cite{thornton2013population, petroff2019fast, platts2019living, zhang2023physics, Petroff:2021wug}.

The DM of FRBs arises from electromagnetic interactions between radio signals and free electrons in the ionized medium distributed along their path from the source to the observer. The total DM can be decomposed into three contributions: from the Milky Way, the intergalactic medium (IGM), and the FRB's host galaxy. Since the IGM generally dominates the observed DM, and because its contribution accumulates over cosmological distances, the DM-redshift relation for FRBs makes them powerful tools for cosmology.
By utilizing FRBs alone or in combination with other cosmological probes,  one can constrain the dark energy equation of state and other cosmological parameters~\cite{Zhou:2014yta, gao2014fast, Walters:2017afr, wei2018cosmology, zhang2020combinations, zhao2020cosmological, qiu2022forecast,Yang:2016zbm}, measure the Hubble constant ($H_0$)~\cite{Li:2017mek, liu2023cosmological, hagstotz2022new, james2022measurement, Fortunato:2024hfm, piratova2025fast}, probe the cosmic reionization history~\cite{zheng2014probing, Caleb:2019apf, Beniamini:2020ane, hashimoto2021revealing}, test the Einstein's equivalence principle~\cite{Wei:2015hwd, Nusser:2016wzr, Tingay:2016tgf}, trace the large scale structure of the universe~\cite{Masui:2015ola}, and place constraints on the photon mass~\cite{Wu:2016brq,Shao:2017tuu} as well as the magnetic fields in the IGM~\cite{Akahori:2016ami}.

Another key application of FRBs  lies in addressing the long-standing ``missing" baryon problem.  The ionized baryons believed to reside in the diffuse IGM are difficult to detect directly, but the FRB DMs offer a promising observational tracer~\cite{Deng_2014, Ravi:2018ose, Munoz:2018mll, li2019cosmology, li2020cosmology, walters2019probing, wei2019constraining}. Recently, 
an optimized method~\cite{macquart2020census}  was proposed to determine the cosmic baryon density from a small sample of five localized FRBs, where the DM is partitioned into two parts: contributions from within the Milky Way and from extragalactic sources (including the IGM and the host galaxy). By maximizing the joint likelihood  of the probability density function (PDF) of the extragalactic DM,   an estimate of the cosmic baryon density is obtained and the result is consistent with those from cosmic microwave background (CMB) and Big Bang nucleosynthesis (BBN) observations~\cite{macquart2020census}. 

This method has since been widely adopted for a variety of cosmological applications, including  determinations of the Hubble constant~\cite{Zhao:2022yiv, wei2023investigating, wu20228, gao2024measuring, wang2025probing, kalita2024fast, gao2024measurement}, measurements of the cosmic baryon density~\cite{yang2022finding},  assessments of the baryon mass fraction in  the IGM~\cite{lin2023probing,wang20238}, exploration of the dark energy equation of state~\cite{wang2025probing}, estimates of the kinematic parameters of the universe~\cite{Fortunato:2023deh,gao2024measurement}, and  constraints on fundamental physics  such as the photon mass and other constants~\cite{lin2023revised,Wang:2023fnn,kalita2024constraining}.

However, in this paper, we identify a critical factor that was omitted in   the PDF for the extragalactic DM in this optimized method. Using simulated FRB datasets, we demonstrate that neglecting this factor results in a statistically significant bias, specifically, the inferred cosmic baryon density deviates from the true (input) value by more than 1$\sigma$ confidence level.  This result highlights the essential role of the missing factor in cosmological applications of FRBs. Ignoring it can lead to erroneous estimates of cosmological parameters, thereby compromising the robustness of conclusions drawn from FRB observations.


\section{Methodology}
\label{secmath}
For an FRB signal  originating outside the Milky Way, the observed dispersion measure (DM) can be decomposed into four distinct contributions:
\begin{eqnarray}
    \mathrm{DM}_\mathrm{obs}(z) = \mathrm{DM}_\mathrm{ISM} + \mathrm{DM}_\mathrm{halo}+ \mathrm{DM}_\mathrm{IGM}(z) + \mathrm{DM}_\mathrm{host}(z),
    \label{eq:1}
\end{eqnarray} 
where $\mathrm{DM}_\mathrm{ISM}$, $\mathrm{DM}_\mathrm{halo}$, 
$\mathrm{DM}_\mathrm{IGM}$ and $\mathrm{DM}_\mathrm{host}$  represent contributions from the Milky Way interstellar medium (ISM), the Milky Way halo, the IGM, and the FRB's host galaxy, respectively.

Since $\mathrm{DM}_\mathrm{ISM}$ and  $\mathrm{DM}_\mathrm{halo}$ originate within  the Milky Way, 
Macquart et al.~\cite{macquart2020census}  introduced the parameter $\mathrm{DM_{ext}}$, defined as 
\begin{eqnarray}\mathrm{DM_{ext}\equiv DM_{obs}-DM_{ISM}-DM_{halo}=DM_{IGM}+DM_{host}},
 \end{eqnarray}   to encapsulate extragalactic contributions to the observed DM. They then developed an optimized approach to constrain cosmological parameters by maximizing the joint likelihood function:
 \begin{equation}\label{EqL}
    \mathcal{L}=\prod_{i=1}^{n}P_{i}(\mathrm{DM_{ext,i}}) ,
\end{equation}
where $P_i(\mathrm{DM_{ext,i}}) $
denotes  the PDF of the extragalactic DM contribution for the $i$-th FRB. 
Using  the general formula for the PDF of the sum of two independent random variables,  $z = x + y$:     
\begin{eqnarray}
P(z)=\int_{-\infty}^{\infty}P_x(x)P_y\left(z-x\right)\mathrm{d}x,
\end{eqnarray}
 one obtains the PDF of $\mathrm{DM_{ext}}$ as: 
\begin{eqnarray}\label{EqPDF1}
      P(\mathrm{DM_{ext}}) = \int_0^\mathrm{DM_{ext}}P_{\mathrm{host}}(\mathrm{DM_{host}})\times  P_{\mathrm{IGM}}(\mathrm{DM_{ext}-DM_{host}}) d \mathrm{DM_{host}}
\end{eqnarray}
with $\mathrm{DM_{ext}-DM_{host}=DM_{IGM}}$.

Due to substantial fluctuations in the electron distribution within the IGM, the actual value of $\mathrm{DM_{IGM}}$ fluctuates significantly around its mean value $\langle\mathrm{DM_{IGM}}\rangle$.   Numerical simulations of the IGM yield an analytic expression for the PDF of  $\mathrm{DM_{IGM}}$~\cite{mcquinn2013locating}, which can be accurately approximated by the form~\cite{macquart2020census,zhang2021intergalactic}
\begin{equation}
    P_{\mathrm{cosmic}}(\Delta)=A\Delta^{-\beta}\exp\left[-\dfrac{(\Delta^{-\alpha}-C_0)^2}{2\alpha^2\sigma_{\mathrm{IGM}}^2}\right],\quad\Delta>0
    \label{eq5}
\end{equation}
with 
\begin{eqnarray} \mathrm{\Delta \equiv  \frac{DM_{IGM}} {\langle{DM_{IGM}}\rangle} = \frac{DM_{ext}-DM_{host}}{\langle{DM_{IGM}}\rangle}}. 
\end{eqnarray} 
Here $\beta$ and $\alpha$  are constants,   parameters $C_0$ and $\sigma_{\mathrm{IGM}}$ are functions of redshift,  and parameter $A$ is a   normalization factor with normalization over $\Delta$ applied  in \cite{macquart2020census}. 

Macquart et al.~\cite{macquart2020census} obtained the PDF of $\mathrm{DM_{ext}}$ by directly replacing $P_\mathrm{IGM}(\mathrm{DM_{ext}}-\mathrm{DM_{host}})$ in Eq.~(\ref{EqPDF1}) with $P_\mathrm{cosmic}(\Delta)$, yielding
\begin{eqnarray}\label{EqPDF2}
P(\mathrm{DM_{ext}}) &=& \int_0^{\mathrm{DM_{ext}}} P_{\mathrm{host}}(\mathrm{DM_{host}})\times P_{\mathrm{cosmic}}\left(\frac{\mathrm{DM_{ext}}-\mathrm{DM_{host}}}{\langle \mathrm{DM_{IGM}}\rangle}\right) d\mathrm{DM_{host}} , .
\end{eqnarray}
However, $P(\mathrm{DM_{ext}})$ as defined in Ref.~\cite{macquart2020census} is not properly normalized.\footnote{Implementation details can be found at \url{https://github.com/FRBs/FRB}.} Despite this, the unnormalized PDF was subsequently substituted into Eq.~(\ref{EqL}) to constrain cosmological parameters.

However, for the PDF of the product of two random variables,  $z = xy$,   the correct mathematical formulation is~  \cite{Rohatgi}
\begin{equation}
    P(z) = \int_{-\infty}^{\infty}\dfrac{1}{\left|x\right|}\times P_x(x) \times P_y\left(\dfrac{z}{x}\right){d}x\;.
\end{equation}
Thus, the PDF of $\mathrm{DM_{IGM}}$ should correctly be expressed as:
\begin{eqnarray}
 P\mathrm{_{IGM}(DM_{ext}-DM_{host})} &=&P\mathrm{_{IGM}(\Delta\times\langle DM_{IGM}\rangle )}\\ \nonumber
    &=&\frac{1}{\mathrm{\langle DM_{IGM}\rangle}}P\mathrm{_{cosmic}\left( \frac{\mathrm{DM_{ext}-DM_{host}}}{\mathrm{\langle DM_{IGM}\rangle}}\right)}.
\end{eqnarray}
Therefore, the correct PDF for $\mathrm{DM_{ext}}$ should be:
\begin{eqnarray}
      P(\mathrm{DM_{ext}}) &=& \int_0^{\mathrm{DM_{ext}}}\dfrac{1}{\mathrm{\langle DM_{IGM}\rangle}}\times P_{\mathrm{host}}(\mathrm{DM_{host}})\times  P_{\mathrm{cosmic}}\left(\dfrac{\mathrm{DM_{ext}-DM_{host}}}{\mathrm{\langle DM_{IGM}\rangle}}\right)d \mathrm{DM_{host}}. 
      \nonumber\\
      \label{eq8}
\end{eqnarray}
Comparing Eqs.~(\ref{EqPDF2}) and (\ref{eq8}), it is evident that the PDF of $\mathrm{DM_{ext}}$ used by Macquart et al.~\cite{macquart2020census} omits the crucial factor $1/\langle\mathrm{DM_{IGM}}\rangle$~\footnote{ The same omission was also noted recently in~\cite{Zhuge:2025urk}.}.   This factor is actually the Jacobian arising from the transformation between $\Delta$ and $\mathrm{DM_{IGM}}$.   Since $1/\langle\mathrm{DM_{IGM}}\rangle$ explicitly depends on cosmological parameters, it cannot be neglected. Omitting this factor inevitably leads to biased parameter constraints, underscoring its fundamental importance for reliable cosmological analyses with FRBs.

\section{Simulation and results}
\label{sec3}

To quantitatively evaluate the influence of the omitted factor $1/\langle\mathrm{DM_{IGM}}\rangle$ in the PDF of $\mathrm{DM_{ext}}$ on cosmological parameter estimation, we simulate mock FRB datasets and then constrain cosmological parameters using both forms of the PDF given by Eqs.~(\ref{EqPDF2}) and (\ref{eq8}).

We  generate a simulated dataset comprising 100 localized FRB events,  roughly matching the number of localized FRBs currently observed~\cite{wang2025probing}. The redshift distribution of these simulated FRBs follows~\cite{qiang2021effect} 
\begin{equation}
    \mathrm{P_{model}}(z)\propto z\exp{(-z)}
\end{equation}
 with an upper redshift  limit $z=1.5$. 
 
 Since the value $\mathrm{DM_{host}}$ strongly depends on the physical properties of the FRB's host galaxy and surrounding plasma environment,   it is difficult to determine precisely. Observations indicate significant variations in the host galaxy DMs among FRB events~\cite{Marcote:2020ljw, Tendulkar:2017vuq, Bannister:2019iju}. Fortunately, the IllustrisTNG simulation  has demonstrated  that the distribution of $\mathrm{DM_{host}}$ is   well-described  by a  log-normal distribution~\cite{zhang2020dispersion}:
\begin{eqnarray}
    P_\mathrm{host}(\mathrm{DM_{host}})=\dfrac{1}{\sqrt{2\pi}\mathrm{DM_{host}\sigma_{host}}}\times \exp\mathrm{\left[-\dfrac{(\ln{DM_{host}}-\mu_{host})^2}{2\sigma_{host}^2}\right]},
    \label{eq3}
\end{eqnarray}
where the parameters $\mathrm{\mu_{host}}$ and $\mathrm{\sigma_{host}}$, obtained by using the IllustrisTNG simulations at discrete redshifts within the interval $z\in [0.1,1.5]$,  are listed in   Table 3
of~\cite{zhang2020dispersion}. We apply cubic spline interpolation to estimate these parameters at the redshifts of the simulated data.

The simulated value of  $\mathrm{DM_{IGM}}$ can be calculated via the relation $\mathrm{DM_{IGM}}=\Delta\times\mathrm{\langle{DM_{IGM}}\rangle}$ with $\Delta$ satisfying the distribution given by  Eq.~(\ref{eq5}). Thus,  we first generate $\Delta$ from Eq.~(\ref{eq5}). Parameters $A$, $C_0$ and $\sigma_{\mathrm{IGM}}$ from Eq.~(\ref{eq5})  have been computed through the IllustrisTNG simulations for several redshift points in the range $z\in[0.1,9]$ (Table 1 in Ref.~\cite{zhang2021intergalactic}),  while parameters $\alpha$ and $\beta$ are fixed at $\alpha=\beta=3$~\cite{macquart2020census}.  We again use   cubic spline interpolation to obtain   $A$, $C_0$ and $\sigma_{\mathrm{IGM}}$  at the redshifts of the mock data.   
The mean intergalactic DM, $\mathrm{\langle{DM_{IGM}}\rangle}$ ,  is calculated using the relation~\cite{ioka2003cosmic, inoue2004probing}:
\begin{eqnarray}
   \mathrm{\langle{DM_{IGM}}\rangle}&=&\frac{3 \Omega_\mathrm{b}  {H_0}f_\mathrm{{IGM}}}{8\pi G m_p}
   \int_0^z\frac{(1+z')\chi_e(z')}{\sqrt{\Omega_\mathrm{m0}(1+z')^3+\Omega_{\Lambda 0}}}{d}z'
  \nonumber \\
   &=&\frac{3 \mathrm{\Omega_b h_{70}}\cdot70(\mathrm{km/(s\cdot Mpc))} f_\mathrm{{IGM}}}{8\pi  G m_p }\times \int_0^z\frac{(1+z')\chi_e(z')}{\sqrt{\Omega_\mathrm{m0}(1+z')^3+\Omega_{\Lambda 0}}} {d}z',
   \label{eq13}
\end{eqnarray}
assuming a spatially flat $\Lambda$CDM cosmological model.
Here,  $\Omega_\mathrm{b}$ is the baryon mass fraction in our universe, $\mathrm{h_{70}}=\frac{ {H_0}}{\mathrm{70km/(s\cdot Mpc)}}$ is the dimensionless Hubble constant with $H_0$ being the Hubble constant, $f_\mathrm{{IGM}}$ is the fraction of baryon mass in the IGM, $m_p$ is the proton mass, and $\Omega_\mathrm{m0}$ and $\Omega_{\Lambda0}$, which satisfy $\Omega_\mathrm{m0}+\Omega_{\Lambda0}=1$,  are the present density parameters for pressureless matter  and cosmological constant dark energy, respectively.  The free electron number fraction per baryon, $\chi_e(z)$, is defined as $\chi_e(z)=\frac{3}{4}\chi_{e,H}(z)+\frac{1}{8}\chi_{e,He}(z)$, where $\chi_{e,H}(z)$ and $\chi_{e,He}(z)$ are the ionization fractions of hydrogen and helium, respectively.   At redshifts $z<3$, 
 both of hydrogen and helium are completely ionized,  thus  $\chi_{e,H}(z)=\chi_{e,He}(z)=1$, which means $\chi_e=\frac{7}{8}$.   In the simulation, we set parameters  $\Omega_\mathrm{m0}=0.315$ and $\Omega_\mathrm{b} \mathrm{h_{70}}=0.0474$  from the Planck 2018 results~\cite{aghanim2020planck},  and  set $f_\mathrm{{IGM}}=0.83$~\cite{Fukugita:1997bi} to generate $\mathrm{\langle{DM_{IGM}}\rangle}$.  Finally,  by multiplying $\Delta$ by $\mathrm{\langle{DM_{IGM}}\rangle}$, we obtain  the simulated vaules of  $\mathrm{DM_{IGM}}$.

\begin{figure}
	\includegraphics[width=0.45\columnwidth]{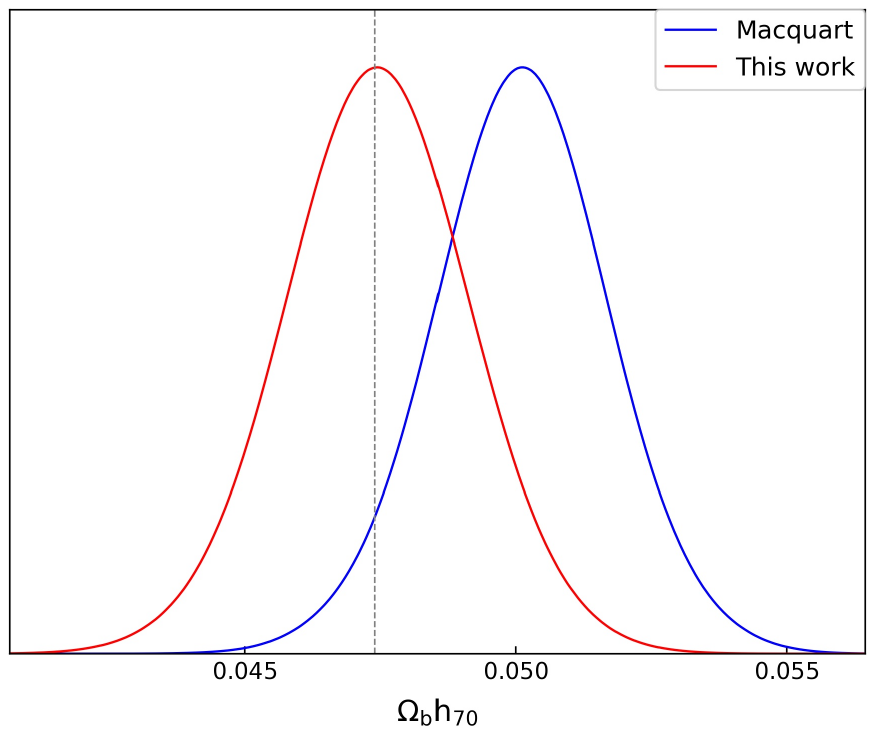}
    \caption{1D marginalized posterior distributions for parameter ${\mathrm{\Omega_b h_{70}}}$ from mock data. The blue and red lines denote consequences from the PDFs of $\mathrm{DM_{ext}}$ given in Eqs.~(\ref{EqPDF2}) and (\ref{eq8}), respectively.
    The gray dashed line represents the fiducial value $\Omega_\mathrm{b}\mathrm{h_{70}}$ = 0.0474 used in simulation. }
    \label{Fig1}
\end{figure}

We next constrain the parameter $\mathrm{\Omega_b h_{70}}$ from the simulated data using the emcee Python package for Markov Chain Monte Carlo (MCMC) sampling~\cite{foreman2013emcee}.  We impose a uniform prior of $0.015\leq\mathrm{\Omega_b h_{70}}\leq0.095$. To minimize statistical fluctuations arising from a single realization, we repeat the simulation and constraint process 100 times, with the combined results presented  in Fig.~\ref{Fig1}. The blue and red lines denote  constraints using the PDFs of $\mathrm{DM_{ext}}$ given in Eqs.~(\ref{EqPDF2}) and (\ref{eq8}), respectively, while gray dashed line represents  the fiducial value   $\mathrm{\Omega_b h_{70}}=0.0474$  used in the simulations. 

Using the PDF derived by Macquart et al.~\cite{macquart2020census}, we obtain $\mathrm{\Omega_b h_{70}}=0.0501\pm{0.0016}$ (at $1\sigma$ confidence level), which deviates from the true simulated value at a significance exceeding $1\sigma$ (approximately $1.6\sigma$ deviation). Conversely, employing our corrected PDF (Eq.(\ref{eq8})), we recover $\mathrm{\Omega_b h_{70}}=0.0475\pm{0.0017}$, fully consistent with the fiducial value. Thus, omitting the $1/\langle\mathrm{DM_{IGM}}\rangle$ factor clearly biases cosmological constraints, confirming the necessity of using Eq.~(\ref{eq8}) for accurate cosmological inference with FRBs.


\section{determine $\mathrm{\Omega_b h_{70}}$ from real data of FRBs}
\label{sec:88}

We now use a recent compilation of 92 localized FRBs~\cite{wang2025probing} to determine the cosmic baryon density.   Following the approach in~\cite{wang2025probing},  we reselect these 92 data points  by using the condition $\mathrm{DM_{obs}-DM_{ISM}-DM_{halo}>80}$ pc cm$^{-3}$, which removes  4 FRBs and  leaves 88 data points for our analysis. Values of $\mathrm{DM_{ISM}}$ are achieved by using the YMW16 model~\cite{yao2017new}. All parameters in $P_\mathrm{host}(\mathrm{DM_{host}})$  and $P_\mathrm{cosmic}(\Delta)$ remain identical to those used in our previous simulations.  We adopt a flat $\Lambda$CDM cosmological model with fixed matter density $\Omega_{m0}=0.315$ and assume the baryon fraction in the IGM to be $f_\mathrm{IGM}=0.83$.  Since the precise contribution of our Galactic halo, $\mathrm{DM_{halo}}$, is uncertain and estimated to range between $50$ and $80\,\mathrm{pc\, cm^{-3}}$~\cite{prochaska2019probing}, we treat it as a free parameter with a Gaussian prior $\mathcal{N}(65,15^2)$ (pc cm$^{-3}$) constrained to the range $50\sim80\,\mathrm{pc\,cm^{-3}}$.

\begin{figure}
	\includegraphics[width=0.55\columnwidth]{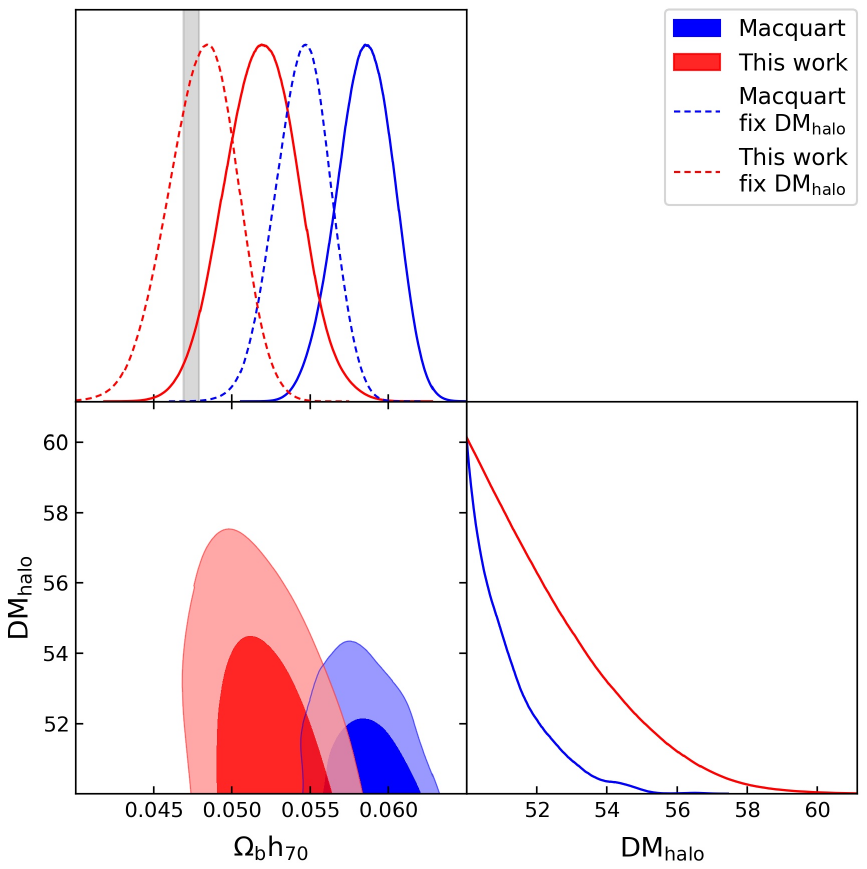}
    \caption{1D marginalized posterior distributions and 2D 1-2$\sigma$ contour regions for parameters ${\mathrm{\Omega_b h_{70}}}$ and $\mathrm{DM_{halo}}$ from 88 localized FRB data. The blue and red lines denote consequences from the PDFs of $\mathrm{DM_{ext}}$ given in Eqs.~(\ref{EqPDF2}) and (\ref{eq8}), respectively.  Dashed lines denotes the constraints on $\mathrm{\Omega_b h_{70}}$ after fixing $\mathrm{DM_{halo}}=60$ pc cm$^{-3}$.  The gray band  represents  $\Omega_b\mathrm{h_{70}}= 0.0474\pm0.0005$  from the Planck 2018 CMB data~\cite{aghanim2020planck}.}
    \label{Fig2}
\end{figure}

Applying the MCMC method with a uniform prior of $0.015<\mathrm{\Omega_b h_{70}}<0.095$, 
we obtain  joint  constraints on $\mathrm{\Omega_b h_{70}} $ and $\mathrm{DM_{halo}}$,  shown in Fig.~\ref{Fig2}. In this Figure,  the solid  red and blue lines represent the results obtained using our corrected method (Eq.(\ref{eq8})) and Macquart et al.'s original method (Eq.(\ref{EqPDF2})), respectively.  The gray shaded band denotes  $\mathrm{\Omega_b h_{70}}= 0.0474\pm0.0005$ from the  Planck 2018 CMB data~\cite{aghanim2020planck}.  Using our corrected PDF, we find $\mathrm{\Omega_b h_{70}}=0.0519\pm{0.0023}$, which aligns with the Planck 2018 result within $2\sigma$. In contrast, the PDF from Macquart et al. yields $\mathrm{\Omega_b h_{70}}=0.0585\pm0.0018$, deviating from the Planck value by $5.8\sigma$.

Notably,  despite adopting a Gaussian prior centered on $\mathrm{DM_{halo}}=65 \,\mathrm{pc\,cm^{-3}}$, the FRB data strongly favor smaller values of $\mathrm{DM_{halo}}$.  Additionally, there is a clear anti-correlation between $\mathrm{\Omega_b h_{70}}$ and $\mathrm{DM_{halo}}$, implying that larger values of $\mathrm{DM_{halo}}$  result in smaller inferred $\mathrm{\Omega_b h_{70}}$. 
This motivates  an additional analysis where we fix $\mathrm{DM_{halo}}=60 \,\mathrm{pc \,cm^{-3}}$.   In this case, we obtain $\mathrm{\Omega_b h_{70}}=0.0545\pm0.0018$ using Macquart et al.'s PDF and $\mathrm{\Omega_b h_{70}}=0.0481^{+0.0023}_{-0.0020}$ using our corrected PDF (indicated by dashed lines in Fig.~\ref{Fig2}). Clearly, the corrected PDF yields a result fully consistent with the Planck 2018 measurement.

\section{Conclusion and discussion}
\label{sec5}

FRBs serve as  powerful cosmological probes. Recently, 
an optimized method was proposed for measuring the cosmic baryon density using localized FRBs,  based on maximizing the joint likelihood function of the extragalactic dispersion measure, $\mathrm{DM_{ext}}$. However,   this method  omits a crucial facto in the probability density function (PDF) for $\mathrm{DM_{ext}}$.

Using simulated FRB datasets, we demonstrate that neglecting this factor leads to biased constraints, with the inferred cosmological parameter deviating from the true simulation input by more than $1\sigma$. In contrast, incorporating the missing factor yields constraints fully consistent with the true simulation input values. Furthermore, applying the corrected PDF to a dataset of 88 real localized FRBs, we find that the inferred cosmic baryon density aligns well with the Planck 2018 CMB measurement,  whereas the uncorrected method results in significant disagreement. We therefore conclude that for reliable cosmological analyses involving FRBs, the corrected PDF (Eq.~(\ref{eq8})) must be used.


\begin{acknowledgments}
This work was supported by the National Natural Science Foundation of China (Grants  No.~12275080 and No.~12075084), the Major basic research project of Hunan Province (Grant No.~2024JC0001), and the Innovative Research Group of Hunan Province (Grant No.~2024JJ1006).
\end{acknowledgments}

\bibliography{ref}

\begin{thebibliography}{69}%
\makeatletter
\providecommand \@ifxundefined [1]{%
 \@ifx{#1\undefined}
}%
\providecommand \@ifnum [1]{%
 \ifnum #1\expandafter \@firstoftwo
 \else \expandafter \@secondoftwo
 \fi
}%
\providecommand \@ifx [1]{%
 \ifx #1\expandafter \@firstoftwo
 \else \expandafter \@secondoftwo
 \fi
}%
\providecommand \natexlab [1]{#1}%
\providecommand \enquote  [1]{``#1''}%
\providecommand \bibnamefont  [1]{#1}%
\providecommand \bibfnamefont [1]{#1}%
\providecommand \citenamefont [1]{#1}%
\providecommand \href@noop [0]{\@secondoftwo}%
\providecommand \href [0]{\begingroup \@sanitize@url \@href}%
\providecommand \@href[1]{\@@startlink{#1}\@@href}%
\providecommand \@@href[1]{\endgroup#1\@@endlink}%
\providecommand \@sanitize@url [0]{\catcode `\\12\catcode `\$12\catcode
  `\&12\catcode `\#12\catcode `\^12\catcode `\_12\catcode `\%12\relax}%
\providecommand \@@startlink[1]{}%
\providecommand \@@endlink[0]{}%
\providecommand \url  [0]{\begingroup\@sanitize@url \@url }%
\providecommand \@url [1]{\endgroup\@href {#1}{\urlprefix }}%
\providecommand \urlprefix  [0]{URL }%
\providecommand \Eprint [0]{\href }%
\providecommand \doibase [0]{http://dx.doi.org/}%
\providecommand \selectlanguage [0]{\@gobble}%
\providecommand \bibinfo  [0]{\@secondoftwo}%
\providecommand \bibfield  [0]{\@secondoftwo}%
\providecommand \translation [1]{[#1]}%
\providecommand \BibitemOpen [0]{}%
\providecommand \bibitemStop [0]{}%
\providecommand \bibitemNoStop [0]{.\EOS\space}%
\providecommand \EOS [0]{\spacefactor3000\relax}%
\providecommand \BibitemShut  [1]{\csname bibitem#1\endcsname}%
\let\auto@bib@innerbib\@empty
\bibitem [{\citenamefont {Lorimer}\ \emph {et~al.}(2007)\citenamefont
  {Lorimer}, \citenamefont {Bailes}, \citenamefont {McLaughlin}, \citenamefont
  {Narkevic},\ and\ \citenamefont {Crawford}}]{lorimer2007bright}%
  \BibitemOpen
  \bibfield  {author} {\bibinfo {author} {\bibfnamefont {D.~R}\ \bibnamefont
  {Lorimer}}, \bibinfo {author} {\bibfnamefont {M.}~\bibnamefont {Bailes}},
  \bibinfo {author} {\bibfnamefont {M.~A.}\ \bibnamefont {McLaughlin}},
  \bibinfo {author} {\bibfnamefont {D.~J}\ \bibnamefont {Narkevic}}, \ and\
  \bibinfo {author} {\bibfnamefont {F.}~\bibnamefont {Crawford}},\ }\bibfield
  {title} {\enquote {\bibinfo {title} {A bright millisecond radio burst of
  extragalactic origin},}\ }\href {\doibase
  https://doi.org/10.1126/science.1147532} {\bibfield  {journal} {\bibinfo
  {journal} {Science}\ }\textbf {\bibinfo {volume} {318}},\ \bibinfo {pages}
  {777--780} (\bibinfo {year} {2007})}\BibitemShut {NoStop}%
\bibitem [{\citenamefont {Thornton}\ \emph {et~al.}(2013)\citenamefont
  {Thornton}, \citenamefont {Stappers}, \citenamefont {Bailes}, \citenamefont
  {Barsdell}, \citenamefont {Bates}, \citenamefont {Bhat}, \citenamefont
  {Burgay}, \citenamefont {Burke-Spolaor}, \citenamefont {Champion},
  \citenamefont {Coster} \emph {et~al.}}]{thornton2013population}%
  \BibitemOpen
  \bibfield  {author} {\bibinfo {author} {\bibfnamefont {D.}~\bibnamefont
  {Thornton}}, \bibinfo {author} {\bibfnamefont {B.}~\bibnamefont {Stappers}},
  \bibinfo {author} {\bibfnamefont {M.}~\bibnamefont {Bailes}}, \bibinfo
  {author} {\bibfnamefont {B.}~\bibnamefont {Barsdell}}, \bibinfo {author}
  {\bibfnamefont {S.}~\bibnamefont {Bates}}, \bibinfo {author} {\bibfnamefont
  {N.}~\bibnamefont {Bhat}}, \bibinfo {author} {\bibfnamefont {M.}~\bibnamefont
  {Burgay}}, \bibinfo {author} {\bibfnamefont {S.}~\bibnamefont
  {Burke-Spolaor}}, \bibinfo {author} {\bibfnamefont {D.}~\bibnamefont
  {Champion}}, \bibinfo {author} {\bibfnamefont {P.}~\bibnamefont {Coster}},
  \emph {et~al.},\ }\bibfield  {title} {\enquote {\bibinfo {title} {A
  population of fast radio bursts at cosmological distances},}\ }\href
  {\doibase https://doi.org/10.1126/science.1236789} {\bibfield  {journal}
  {\bibinfo  {journal} {Science}\ }\textbf {\bibinfo {volume} {341}},\ \bibinfo
  {pages} {53--56} (\bibinfo {year} {2013})}\BibitemShut {NoStop}%
\bibitem [{\citenamefont {Petroff}\ \emph {et~al.}(2019)\citenamefont
  {Petroff}, \citenamefont {Hessels},\ and\ \citenamefont
  {Lorimer}}]{petroff2019fast}%
  \BibitemOpen
  \bibfield  {author} {\bibinfo {author} {\bibfnamefont {E.}~\bibnamefont
  {Petroff}}, \bibinfo {author} {\bibfnamefont {J.}~\bibnamefont {Hessels}}, \
  and\ \bibinfo {author} {\bibfnamefont {D.}~\bibnamefont {Lorimer}},\
  }\bibfield  {title} {\enquote {\bibinfo {title} {Fast radio bursts},}\ }\href
  {\doibase https://doi.org/10.1007/s00159-019-0116-6} {\bibfield  {journal}
  {\bibinfo  {journal} {Astron. Astrophys. Rev.}\ }\textbf {\bibinfo {volume}
  {27}},\ \bibinfo {pages} {4} (\bibinfo {year} {2019})}\BibitemShut {NoStop}%
\bibitem [{\citenamefont {Platts}\ \emph {et~al.}(2019)\citenamefont {Platts},
  \citenamefont {Weltman}, \citenamefont {Walters}, \citenamefont {Tendulkar},
  \citenamefont {Gordin},\ and\ \citenamefont {Kandhai}}]{platts2019living}%
  \BibitemOpen
  \bibfield  {author} {\bibinfo {author} {\bibfnamefont {E.}~\bibnamefont
  {Platts}}, \bibinfo {author} {\bibfnamefont {A.}~\bibnamefont {Weltman}},
  \bibinfo {author} {\bibfnamefont {A.}~\bibnamefont {Walters}}, \bibinfo
  {author} {\bibfnamefont {S.~P.}\ \bibnamefont {Tendulkar}}, \bibinfo {author}
  {\bibfnamefont {J.~E.~B.}\ \bibnamefont {Gordin}}, \ and\ \bibinfo {author}
  {\bibfnamefont {S.}~\bibnamefont {Kandhai}},\ }\bibfield  {title} {\enquote
  {\bibinfo {title} {A living theory catalogue for fast radio bursts},}\ }\href
  {\doibase https://doi.org/10.1016/j.physrep.2019.06.003} {\bibfield
  {journal} {\bibinfo  {journal} {Physics Reports}\ }\textbf {\bibinfo {volume}
  {821}},\ \bibinfo {pages} {1--27} (\bibinfo {year} {2019})}\BibitemShut
  {NoStop}%
\bibitem [{\citenamefont {Zhang}(2023)}]{zhang2023physics}%
  \BibitemOpen
  \bibfield  {author} {\bibinfo {author} {\bibfnamefont {B.}~\bibnamefont
  {Zhang}},\ }\bibfield  {title} {\enquote {\bibinfo {title} {The physics of
  fast radio bursts},}\ }\href {\doibase
  https://doi.org/10.1103/RevModPhys.95.035005} {\bibfield  {journal} {\bibinfo
   {journal} {Rev. Mod. Phys.}\ }\textbf {\bibinfo {volume} {95}},\ \bibinfo
  {pages} {035005} (\bibinfo {year} {2023})}\BibitemShut {NoStop}%
\bibitem [{\citenamefont {Petroff}\ \emph {et~al.}(2022)\citenamefont
  {Petroff}, \citenamefont {Hessels},\ and\ \citenamefont
  {Lorimer}}]{Petroff:2021wug}%
  \BibitemOpen
  \bibfield  {author} {\bibinfo {author} {\bibfnamefont {E.}~\bibnamefont
  {Petroff}}, \bibinfo {author} {\bibfnamefont {J.~W.~T.}\ \bibnamefont
  {Hessels}}, \ and\ \bibinfo {author} {\bibfnamefont {D.~R.}\ \bibnamefont
  {Lorimer}},\ }\bibfield  {title} {\enquote {\bibinfo {title} {{Fast radio
  bursts at the dawn of the 2020s}},}\ }\href {\doibase
  10.1007/s00159-022-00139-w} {\bibfield  {journal} {\bibinfo  {journal}
  {Astron. Astrophys. Rev.}\ }\textbf {\bibinfo {volume} {30}},\ \bibinfo
  {pages} {2} (\bibinfo {year} {2022})}\BibitemShut {NoStop}%
\bibitem [{\citenamefont {Zhou}\ \emph {et~al.}(2014)\citenamefont {Zhou},
  \citenamefont {Li}, \citenamefont {Wang}, \citenamefont {Fan},\ and\
  \citenamefont {Wei}}]{Zhou:2014yta}%
  \BibitemOpen
  \bibfield  {author} {\bibinfo {author} {\bibfnamefont {B.}~\bibnamefont
  {Zhou}}, \bibinfo {author} {\bibfnamefont {X.}~\bibnamefont {Li}}, \bibinfo
  {author} {\bibfnamefont {T.}~\bibnamefont {Wang}}, \bibinfo {author}
  {\bibfnamefont {Y.~Z.}\ \bibnamefont {Fan}}, \ and\ \bibinfo {author}
  {\bibfnamefont {D.~M.}\ \bibnamefont {Wei}},\ }\bibfield  {title} {\enquote
  {\bibinfo {title} {{Fast radio bursts as a cosmic probe?}}}\ }\href {\doibase
  10.1103/PhysRevD.89.107303} {\bibfield  {journal} {\bibinfo  {journal} {Phys.
  Rev. D}\ }\textbf {\bibinfo {volume} {89}},\ \bibinfo {pages} {107303}
  (\bibinfo {year} {2014})}\BibitemShut {NoStop}%
\bibitem [{\citenamefont {Gao}\ \emph {et~al.}(2014)\citenamefont {Gao},
  \citenamefont {Li},\ and\ \citenamefont {Zhang}}]{gao2014fast}%
  \BibitemOpen
  \bibfield  {author} {\bibinfo {author} {\bibfnamefont {H.}~\bibnamefont
  {Gao}}, \bibinfo {author} {\bibfnamefont {Z.}~\bibnamefont {Li}}, \ and\
  \bibinfo {author} {\bibfnamefont {B.}~\bibnamefont {Zhang}},\ }\bibfield
  {title} {\enquote {\bibinfo {title} {Fast radio burst/gamma-ray burst
  cosmography},}\ }\href {\doibase https://doi.org/10.1088/0004-637X/788/2/189}
  {\bibfield  {journal} {\bibinfo  {journal} {Astrophys. J.}\ }\textbf
  {\bibinfo {volume} {788}},\ \bibinfo {pages} {189} (\bibinfo {year}
  {2014})}\BibitemShut {NoStop}%
\bibitem [{\citenamefont {Walters}\ \emph {et~al.}(2018)\citenamefont
  {Walters}, \citenamefont {Weltman}, \citenamefont {Gaensler}, \citenamefont
  {Ma},\ and\ \citenamefont {Witzemann}}]{Walters:2017afr}%
  \BibitemOpen
  \bibfield  {author} {\bibinfo {author} {\bibfnamefont {A.}~\bibnamefont
  {Walters}}, \bibinfo {author} {\bibfnamefont {A.}~\bibnamefont {Weltman}},
  \bibinfo {author} {\bibfnamefont {B.~M.}\ \bibnamefont {Gaensler}}, \bibinfo
  {author} {\bibfnamefont {Y.~Z.}\ \bibnamefont {Ma}}, \ and\ \bibinfo {author}
  {\bibfnamefont {A.}~\bibnamefont {Witzemann}},\ }\bibfield  {title} {\enquote
  {\bibinfo {title} {{Future Cosmological Constraints from Fast Radio
  Bursts}},}\ }\href {\doibase 10.3847/1538-4357/aaaf6b} {\bibfield  {journal}
  {\bibinfo  {journal} {Astrophys. J.}\ }\textbf {\bibinfo {volume} {856}},\
  \bibinfo {pages} {65} (\bibinfo {year} {2018})}\BibitemShut {NoStop}%
\bibitem [{\citenamefont {Wei}\ \emph {et~al.}(2018)\citenamefont {Wei},
  \citenamefont {Wu},\ and\ \citenamefont {Gao}}]{wei2018cosmology}%
  \BibitemOpen
  \bibfield  {author} {\bibinfo {author} {\bibfnamefont {J.~J.}\ \bibnamefont
  {Wei}}, \bibinfo {author} {\bibfnamefont {X.~F.}\ \bibnamefont {Wu}}, \ and\
  \bibinfo {author} {\bibfnamefont {H.}~\bibnamefont {Gao}},\ }\bibfield
  {title} {\enquote {\bibinfo {title} {Cosmology with gravitational wave/fast
  radio burst associations},}\ }\href {\doibase
  https://doi.org/10.3847/2041-8213/aac8e2} {\bibfield  {journal} {\bibinfo
  {journal} {Astrophys. J. Lett.}\ }\textbf {\bibinfo {volume} {860}},\
  \bibinfo {pages} {L7} (\bibinfo {year} {2018})}\BibitemShut {NoStop}%
\bibitem [{\citenamefont {Zhang}\ and\ \citenamefont
  {Li}(2020)}]{zhang2020combinations}%
  \BibitemOpen
  \bibfield  {author} {\bibinfo {author} {\bibfnamefont {L.}~\bibnamefont
  {Zhang}}\ and\ \bibinfo {author} {\bibfnamefont {Z.~X.}\ \bibnamefont {Li}},\
  }\bibfield  {title} {\enquote {\bibinfo {title} {Combinations of standard
  pings and standard candles: An effective and hubble constant-free probe of
  dark energy evolution},}\ }\href {\doibase
  https://doi.org/10.3847/1538-4357/abb091} {\bibfield  {journal} {\bibinfo
  {journal} {Astrophys. J.}\ }\textbf {\bibinfo {volume} {901}},\ \bibinfo
  {pages} {130} (\bibinfo {year} {2020})}\BibitemShut {NoStop}%
\bibitem [{\citenamefont {Zhao}\ \emph {et~al.}(2020)\citenamefont {Zhao},
  \citenamefont {Li}, \citenamefont {Qi}, \citenamefont {Gao}, \citenamefont
  {Zhang},\ and\ \citenamefont {Zhang}}]{zhao2020cosmological}%
  \BibitemOpen
  \bibfield  {author} {\bibinfo {author} {\bibfnamefont {Z.~W.}\ \bibnamefont
  {Zhao}}, \bibinfo {author} {\bibfnamefont {Z.~X.}\ \bibnamefont {Li}},
  \bibinfo {author} {\bibfnamefont {J.~Z.}\ \bibnamefont {Qi}}, \bibinfo
  {author} {\bibfnamefont {H.}~\bibnamefont {Gao}}, \bibinfo {author}
  {\bibfnamefont {J.~F.}\ \bibnamefont {Zhang}}, \ and\ \bibinfo {author}
  {\bibfnamefont {X.}~\bibnamefont {Zhang}},\ }\bibfield  {title} {\enquote
  {\bibinfo {title} {Cosmological parameter estimation for dynamical dark
  energy models with future fast radio burst observations},}\ }\href {\doibase
  https://doi.org/10.3847/1538-4357/abb8ce} {\bibfield  {journal} {\bibinfo
  {journal} {Astrophys. J.}\ }\textbf {\bibinfo {volume} {903}},\ \bibinfo
  {pages} {83} (\bibinfo {year} {2020})}\BibitemShut {NoStop}%
\bibitem [{\citenamefont {Qiu}\ \emph {et~al.}(2022)\citenamefont {Qiu},
  \citenamefont {Zhao}, \citenamefont {Wang}, \citenamefont {Zhang},\ and\
  \citenamefont {Zhang}}]{qiu2022forecast}%
  \BibitemOpen
  \bibfield  {author} {\bibinfo {author} {\bibfnamefont {X.~W.}\ \bibnamefont
  {Qiu}}, \bibinfo {author} {\bibfnamefont {Z.~W.}\ \bibnamefont {Zhao}},
  \bibinfo {author} {\bibfnamefont {L.~F.}\ \bibnamefont {Wang}}, \bibinfo
  {author} {\bibfnamefont {J.~F.}\ \bibnamefont {Zhang}}, \ and\ \bibinfo
  {author} {\bibfnamefont {X.}~\bibnamefont {Zhang}},\ }\bibfield  {title}
  {\enquote {\bibinfo {title} {A forecast of using fast radio burst
  observations to constrain holographic dark energy},}\ }\href {\doibase
  https://doi.org/10.1088/1475-7516/2022/02/006} {\bibfield  {journal}
  {\bibinfo  {journal} {JCAP}\ }\textbf {\bibinfo {volume} {02}},\ \bibinfo
  {pages} {006} (\bibinfo {year} {2022})}\BibitemShut {NoStop}%
\bibitem [{\citenamefont {Yang}\ and\ \citenamefont
  {Zhang}(2016)}]{Yang:2016zbm}%
  \BibitemOpen
  \bibfield  {author} {\bibinfo {author} {\bibfnamefont {Y.~P.}\ \bibnamefont
  {Yang}}\ and\ \bibinfo {author} {\bibfnamefont {B.}~\bibnamefont {Zhang}},\
  }\bibfield  {title} {\enquote {\bibinfo {title} {{Extracting host galaxy
  dispersion measure and constraining cosmological parameters using fast radio
  burst data}},}\ }\href {\doibase 10.3847/2041-8205/830/2/L31} {\bibfield
  {journal} {\bibinfo  {journal} {Astrophys. J. Lett.}\ }\textbf {\bibinfo
  {volume} {830}},\ \bibinfo {pages} {L31} (\bibinfo {year}
  {2016})}\BibitemShut {NoStop}%
\bibitem [{\citenamefont {Li}\ \emph {et~al.}(2018)\citenamefont {Li},
  \citenamefont {Gao}, \citenamefont {Ding}, \citenamefont {Wang},\ and\
  \citenamefont {Zhang}}]{Li:2017mek}%
  \BibitemOpen
  \bibfield  {author} {\bibinfo {author} {\bibfnamefont {Z.~X.}\ \bibnamefont
  {Li}}, \bibinfo {author} {\bibfnamefont {H.}~\bibnamefont {Gao}}, \bibinfo
  {author} {\bibfnamefont {X.~H.}\ \bibnamefont {Ding}}, \bibinfo {author}
  {\bibfnamefont {G.~J.}\ \bibnamefont {Wang}}, \ and\ \bibinfo {author}
  {\bibfnamefont {B.}~\bibnamefont {Zhang}},\ }\bibfield  {title} {\enquote
  {\bibinfo {title} {{Strongly lensed repeating fast radio bursts as precision
  probes of the universe}},}\ }\href {\doibase 10.1038/s41467-018-06303-0}
  {\bibfield  {journal} {\bibinfo  {journal} {Nature Commun.}\ }\textbf
  {\bibinfo {volume} {9}},\ \bibinfo {pages} {3833} (\bibinfo {year}
  {2018})}\BibitemShut {NoStop}%
\bibitem [{\citenamefont {Liu}\ \emph {et~al.}(2023)\citenamefont {Liu},
  \citenamefont {Yu},\ and\ \citenamefont {Wu}}]{liu2023cosmological}%
  \BibitemOpen
  \bibfield  {author} {\bibinfo {author} {\bibfnamefont {Y.}~\bibnamefont
  {Liu}}, \bibinfo {author} {\bibfnamefont {H.}~\bibnamefont {Yu}}, \ and\
  \bibinfo {author} {\bibfnamefont {P.}~\bibnamefont {Wu}},\ }\bibfield
  {title} {\enquote {\bibinfo {title} {Cosmological-model-independent
  determination of hubble constant from fast radio bursts and hubble parameter
  measurements},}\ }\href {\doibase 10.3847/2041-8213/acc650} {\bibfield
  {journal} {\bibinfo  {journal} {Astrophys. J. Lett.}\ }\textbf {\bibinfo
  {volume} {946}},\ \bibinfo {pages} {L49} (\bibinfo {year}
  {2023})}\BibitemShut {NoStop}%
\bibitem [{\citenamefont {Hagstotz}\ \emph {et~al.}(2022)\citenamefont
  {Hagstotz}, \citenamefont {Reischke},\ and\ \citenamefont
  {Lilow}}]{hagstotz2022new}%
  \BibitemOpen
  \bibfield  {author} {\bibinfo {author} {\bibfnamefont {S.}~\bibnamefont
  {Hagstotz}}, \bibinfo {author} {\bibfnamefont {R.}~\bibnamefont {Reischke}},
  \ and\ \bibinfo {author} {\bibfnamefont {R.}~\bibnamefont {Lilow}},\
  }\bibfield  {title} {\enquote {\bibinfo {title} {A new measurement of the
  hubble constant using fast radio bursts},}\ }\href {\doibase
  10.1093/mnras/stac077} {\bibfield  {journal} {\bibinfo  {journal} {Mon. Not.
  Roy. Astron. Soc.}\ }\textbf {\bibinfo {volume} {511}},\ \bibinfo {pages}
  {662--667} (\bibinfo {year} {2022})}\BibitemShut {NoStop}%
\bibitem [{\citenamefont {James}\ \emph {et~al.}(2022)\citenamefont {James},
  \citenamefont {Ghosh}, \citenamefont {Prochaska}, \citenamefont {Bannister},
  \citenamefont {Bhandari}, \citenamefont {Day}, \citenamefont {Deller},
  \citenamefont {Glowacki}, \citenamefont {Gordon}, \citenamefont {Heintz}
  \emph {et~al.}}]{james2022measurement}%
  \BibitemOpen
  \bibfield  {author} {\bibinfo {author} {\bibfnamefont {C.~W.}\ \bibnamefont
  {James}}, \bibinfo {author} {\bibfnamefont {E.~M.}\ \bibnamefont {Ghosh}},
  \bibinfo {author} {\bibfnamefont {J.~X.}\ \bibnamefont {Prochaska}}, \bibinfo
  {author} {\bibfnamefont {K.~W.}\ \bibnamefont {Bannister}}, \bibinfo {author}
  {\bibfnamefont {S.}~\bibnamefont {Bhandari}}, \bibinfo {author}
  {\bibfnamefont {C.~K.}\ \bibnamefont {Day}}, \bibinfo {author} {\bibfnamefont
  {A.~T.}\ \bibnamefont {Deller}}, \bibinfo {author} {\bibfnamefont
  {M.}~\bibnamefont {Glowacki}}, \bibinfo {author} {\bibfnamefont {A.~C.}\
  \bibnamefont {Gordon}}, \bibinfo {author} {\bibfnamefont {K.~E.}\
  \bibnamefont {Heintz}},  \emph {et~al.},\ }\bibfield  {title} {\enquote
  {\bibinfo {title} {A measurement of hubble's constant using fast radio
  bursts},}\ }\href {\doibase 10.1093/mnras/stac2524} {\bibfield  {journal}
  {\bibinfo  {journal} {Mon. Not. Roy. Astron. Soc.}\ }\textbf {\bibinfo
  {volume} {516}},\ \bibinfo {pages} {4862--4881} (\bibinfo {year}
  {2022})}\BibitemShut {NoStop}%
\bibitem [{\citenamefont {Fortunato}\ \emph {et~al.}(2025)\citenamefont
  {Fortunato}, \citenamefont {Bacon}, \citenamefont {Hip\'olito-Ricaldi},\ and\
  \citenamefont {Wands}}]{Fortunato:2024hfm}%
  \BibitemOpen
  \bibfield  {author} {\bibinfo {author} {\bibfnamefont {J.~A.~S.}\
  \bibnamefont {Fortunato}}, \bibinfo {author} {\bibfnamefont {D.~J.}\
  \bibnamefont {Bacon}}, \bibinfo {author} {\bibfnamefont {W.~S.}\ \bibnamefont
  {Hip\'olito-Ricaldi}}, \ and\ \bibinfo {author} {\bibfnamefont
  {D.}~\bibnamefont {Wands}},\ }\bibfield  {title} {\enquote {\bibinfo {title}
  {{Fast Radio Bursts and Artificial Neural Networks: a
  cosmological-model-independent estimation of the Hubble constant}},}\ }\href
  {\doibase 10.1088/1475-7516/2025/01/018} {\bibfield  {journal} {\bibinfo
  {journal} {JCAP}\ }\textbf {\bibinfo {volume} {01}},\ \bibinfo {pages} {018}
  (\bibinfo {year} {2025})}\BibitemShut {NoStop}%
\bibitem [{\citenamefont {Piratova-Moreno}\ \emph {et~al.}(2025)\citenamefont
  {Piratova-Moreno}, \citenamefont {Garc{\'\i}a}, \citenamefont
  {Benavides-Gallego},\ and\ \citenamefont {Cabrera}}]{piratova2025fast}%
  \BibitemOpen
  \bibfield  {author} {\bibinfo {author} {\bibfnamefont {E.~F.}\ \bibnamefont
  {Piratova-Moreno}}, \bibinfo {author} {\bibfnamefont {L.}~\bibnamefont
  {Garc{\'\i}a}}, \bibinfo {author} {\bibfnamefont {C.~A.}\ \bibnamefont
  {Benavides-Gallego}}, \ and\ \bibinfo {author} {\bibfnamefont
  {C.}~\bibnamefont {Cabrera}},\ }\bibfield  {title} {\enquote {\bibinfo
  {title} {Fast radio bursts as cosmological proxies: estimating the hubble
  constant},}\ }\href@noop {} {\  (\bibinfo {year} {2025})},\ \Eprint
  {http://arxiv.org/abs/2502.08509} {arXiv:2502.08509 [astro-ph.CO]}
  \BibitemShut {NoStop}%
\bibitem [{\citenamefont {Zheng}\ \emph {et~al.}(2014)\citenamefont {Zheng},
  \citenamefont {Ofek}, \citenamefont {Kulkarni}, \citenamefont {Neill},\ and\
  \citenamefont {Juric}}]{zheng2014probing}%
  \BibitemOpen
  \bibfield  {author} {\bibinfo {author} {\bibfnamefont {Z.}~\bibnamefont
  {Zheng}}, \bibinfo {author} {\bibfnamefont {E.~O.}\ \bibnamefont {Ofek}},
  \bibinfo {author} {\bibfnamefont {S.~R.}\ \bibnamefont {Kulkarni}}, \bibinfo
  {author} {\bibfnamefont {J.~D.}\ \bibnamefont {Neill}}, \ and\ \bibinfo
  {author} {\bibfnamefont {M.}~\bibnamefont {Juric}},\ }\bibfield  {title}
  {\enquote {\bibinfo {title} {Probing the intergalactic medium with fast radio
  bursts},}\ }\href {\doibase 10.1088/0004-637X/797/1/71} {\bibfield  {journal}
  {\bibinfo  {journal} {Astrophys. J.}\ }\textbf {\bibinfo {volume} {797}},\
  \bibinfo {pages} {71} (\bibinfo {year} {2014})}\BibitemShut {NoStop}%
\bibitem [{\citenamefont {Caleb}\ \emph {et~al.}(2019)\citenamefont {Caleb},
  \citenamefont {Flynn},\ and\ \citenamefont {Stappers}}]{Caleb:2019apf}%
  \BibitemOpen
  \bibfield  {author} {\bibinfo {author} {\bibfnamefont {M.}~\bibnamefont
  {Caleb}}, \bibinfo {author} {\bibfnamefont {C.}~\bibnamefont {Flynn}}, \ and\
  \bibinfo {author} {\bibfnamefont {B.}~\bibnamefont {Stappers}},\ }\bibfield
  {title} {\enquote {\bibinfo {title} {{Constraining the era of helium
  reionization using fast radio bursts}},}\ }\href {\doibase
  10.1093/mnras/stz571} {\bibfield  {journal} {\bibinfo  {journal} {Mon. Not.
  Roy. Astron. Soc.}\ }\textbf {\bibinfo {volume} {485}},\ \bibinfo {pages}
  {2281--2286} (\bibinfo {year} {2019})}\BibitemShut {NoStop}%
\bibitem [{\citenamefont {Beniamini}\ \emph {et~al.}(2021)\citenamefont
  {Beniamini}, \citenamefont {Kumar}, \citenamefont {Ma},\ and\ \citenamefont
  {Quataert}}]{Beniamini:2020ane}%
  \BibitemOpen
  \bibfield  {author} {\bibinfo {author} {\bibfnamefont {P.}~\bibnamefont
  {Beniamini}}, \bibinfo {author} {\bibfnamefont {P.}~\bibnamefont {Kumar}},
  \bibinfo {author} {\bibfnamefont {X.}~\bibnamefont {Ma}}, \ and\ \bibinfo
  {author} {\bibfnamefont {E.}~\bibnamefont {Quataert}},\ }\bibfield  {title}
  {\enquote {\bibinfo {title} {{Exploring the epoch of hydrogen reionization
  using FRBs}},}\ }\href {\doibase 10.1093/mnras/stab309} {\bibfield  {journal}
  {\bibinfo  {journal} {Mon. Not. Roy. Astron. Soc.}\ }\textbf {\bibinfo
  {volume} {502}},\ \bibinfo {pages} {5134--5146} (\bibinfo {year}
  {2021})}\BibitemShut {NoStop}%
\bibitem [{\citenamefont {Hashimoto}\ \emph {et~al.}(2021)\citenamefont
  {Hashimoto}, \citenamefont {Goto}, \citenamefont {Lu}, \citenamefont {On},
  \citenamefont {Santos}, \citenamefont {Kim}, \citenamefont {Eser},
  \citenamefont {Ho}, \citenamefont {Hsiao},\ and\ \citenamefont
  {Lin}}]{hashimoto2021revealing}%
  \BibitemOpen
  \bibfield  {author} {\bibinfo {author} {\bibfnamefont {T.}~\bibnamefont
  {Hashimoto}}, \bibinfo {author} {\bibfnamefont {T.}~\bibnamefont {Goto}},
  \bibinfo {author} {\bibfnamefont {T.~Y.}\ \bibnamefont {Lu}}, \bibinfo
  {author} {\bibfnamefont {A.~Y.~L.}\ \bibnamefont {On}}, \bibinfo {author}
  {\bibfnamefont {D.~J.~D.}\ \bibnamefont {Santos}}, \bibinfo {author}
  {\bibfnamefont {S.~J.}\ \bibnamefont {Kim}}, \bibinfo {author} {\bibfnamefont
  {E.~K.}\ \bibnamefont {Eser}}, \bibinfo {author} {\bibfnamefont {S.~C.~C.}\
  \bibnamefont {Ho}}, \bibinfo {author} {\bibfnamefont {T.~Y.}\ \bibnamefont
  {Hsiao}}, \ and\ \bibinfo {author} {\bibfnamefont {L.~Y.}\ \bibnamefont
  {Lin}},\ }\bibfield  {title} {\enquote {\bibinfo {title} {Revealing the
  cosmic reionization history with fast radio bursts in the era of square
  kilometre array},}\ }\href {\doibase 10.1093/mnras/stab186} {\bibfield
  {journal} {\bibinfo  {journal} {Mon. Not. Roy. Astron. Soc.}\ }\textbf
  {\bibinfo {volume} {502}},\ \bibinfo {pages} {2346--2355} (\bibinfo {year}
  {2021})}\BibitemShut {NoStop}%
\bibitem [{\citenamefont {Wei}\ \emph {et~al.}(2015)\citenamefont {Wei},
  \citenamefont {Gao}, \citenamefont {Wu},\ and\ \citenamefont
  {M\'esz\'aros}}]{Wei:2015hwd}%
  \BibitemOpen
  \bibfield  {author} {\bibinfo {author} {\bibfnamefont {J.~J.}\ \bibnamefont
  {Wei}}, \bibinfo {author} {\bibfnamefont {H.}~\bibnamefont {Gao}}, \bibinfo
  {author} {\bibfnamefont {X.~F.}\ \bibnamefont {Wu}}, \ and\ \bibinfo {author}
  {\bibfnamefont {P.}~\bibnamefont {M\'esz\'aros}},\ }\bibfield  {title}
  {\enquote {\bibinfo {title} {{Testing Einstein\textquoteright{}s Equivalence
  Principle With Fast Radio Bursts}},}\ }\href {\doibase
  10.1103/PhysRevLett.115.261101} {\bibfield  {journal} {\bibinfo  {journal}
  {Phys. Rev. Lett.}\ }\textbf {\bibinfo {volume} {115}},\ \bibinfo {pages}
  {261101} (\bibinfo {year} {2015})}\BibitemShut {NoStop}%
\bibitem [{\citenamefont {Nusser}(2016)}]{Nusser:2016wzr}%
  \BibitemOpen
  \bibfield  {author} {\bibinfo {author} {\bibfnamefont {A.}~\bibnamefont
  {Nusser}},\ }\bibfield  {title} {\enquote {\bibinfo {title} {{On Testing the
  Equivalence Principle with Extragalactic Bursts}},}\ }\href {\doibase
  10.3847/2041-8205/821/1/L2} {\bibfield  {journal} {\bibinfo  {journal}
  {Astrophys. J. Lett.}\ }\textbf {\bibinfo {volume} {821}},\ \bibinfo {pages}
  {L2} (\bibinfo {year} {2016})}\BibitemShut {NoStop}%
\bibitem [{\citenamefont {Tingay}\ and\ \citenamefont
  {Kaplan}(2016)}]{Tingay:2016tgf}%
  \BibitemOpen
  \bibfield  {author} {\bibinfo {author} {\bibfnamefont {S.~J.}\ \bibnamefont
  {Tingay}}\ and\ \bibinfo {author} {\bibfnamefont {D.~L.}\ \bibnamefont
  {Kaplan}},\ }\bibfield  {title} {\enquote {\bibinfo {title} {{Limits on
  Einstein\textquoteright{}s Equivalence Principle From the First Localized
  Fast Radio Burst frb 150418}},}\ }\href {\doibase
  10.3847/2041-8205/820/2/L31} {\bibfield  {journal} {\bibinfo  {journal}
  {Astrophys. J. Lett.}\ }\textbf {\bibinfo {volume} {820}},\ \bibinfo {pages}
  {L31} (\bibinfo {year} {2016})}\BibitemShut {NoStop}%
\bibitem [{\citenamefont {Masui}\ and\ \citenamefont
  {Sigurdson}(2015)}]{Masui:2015ola}%
  \BibitemOpen
  \bibfield  {author} {\bibinfo {author} {\bibfnamefont {K.~W.}\ \bibnamefont
  {Masui}}\ and\ \bibinfo {author} {\bibfnamefont {K.}~\bibnamefont
  {Sigurdson}},\ }\bibfield  {title} {\enquote {\bibinfo {title} {Dispersion
  distance and the matter distribution of the universe in dispersion space},}\
  }\href {\doibase 10.1103/PhysRevLett.115.121301} {\bibfield  {journal}
  {\bibinfo  {journal} {Phys. Rev. Lett.}\ }\textbf {\bibinfo {volume} {115}},\
  \bibinfo {pages} {121301} (\bibinfo {year} {2015})}\BibitemShut {NoStop}%
\bibitem [{\citenamefont {Wu}\ \emph {et~al.}(2016)\citenamefont {Wu},
  \citenamefont {Zhang}, \citenamefont {Gao}, \citenamefont {Wei},
  \citenamefont {Zou}, \citenamefont {Lei}, \citenamefont {Zhang},
  \citenamefont {Dai},\ and\ \citenamefont {M\'esz\'aros}}]{Wu:2016brq}%
  \BibitemOpen
  \bibfield  {author} {\bibinfo {author} {\bibfnamefont {X.~F.}\ \bibnamefont
  {Wu}}, \bibinfo {author} {\bibfnamefont {S.~B.}\ \bibnamefont {Zhang}},
  \bibinfo {author} {\bibfnamefont {H.}~\bibnamefont {Gao}}, \bibinfo {author}
  {\bibfnamefont {J.~J.}\ \bibnamefont {Wei}}, \bibinfo {author} {\bibfnamefont
  {Y.~C.}\ \bibnamefont {Zou}}, \bibinfo {author} {\bibfnamefont {W.~H.}\
  \bibnamefont {Lei}}, \bibinfo {author} {\bibfnamefont {B.}~\bibnamefont
  {Zhang}}, \bibinfo {author} {\bibfnamefont {Z.~G.}\ \bibnamefont {Dai}}, \
  and\ \bibinfo {author} {\bibfnamefont {P.}~\bibnamefont {M\'esz\'aros}},\
  }\bibfield  {title} {\enquote {\bibinfo {title} {{Constraints on the Photon
  Mass with Fast Radio Bursts}},}\ }\href {\doibase
  10.3847/2041-8205/822/1/L15} {\bibfield  {journal} {\bibinfo  {journal}
  {Astrophys. J. Lett.}\ }\textbf {\bibinfo {volume} {822}},\ \bibinfo {pages}
  {L15} (\bibinfo {year} {2016})}\BibitemShut {NoStop}%
\bibitem [{\citenamefont {Shao}\ and\ \citenamefont
  {Zhang}(2017)}]{Shao:2017tuu}%
  \BibitemOpen
  \bibfield  {author} {\bibinfo {author} {\bibfnamefont {L.}~\bibnamefont
  {Shao}}\ and\ \bibinfo {author} {\bibfnamefont {B.}~\bibnamefont {Zhang}},\
  }\bibfield  {title} {\enquote {\bibinfo {title} {{Bayesian framework to
  constrain the photon mass with a catalog of fast radio bursts}},}\ }\href
  {\doibase 10.1103/PhysRevD.95.123010} {\bibfield  {journal} {\bibinfo
  {journal} {Phys. Rev. D}\ }\textbf {\bibinfo {volume} {95}},\ \bibinfo
  {pages} {123010} (\bibinfo {year} {2017})}\BibitemShut {NoStop}%
\bibitem [{\citenamefont {Akahori}\ \emph {et~al.}(2016)\citenamefont
  {Akahori}, \citenamefont {Ryu},\ and\ \citenamefont
  {Gaensler}}]{Akahori:2016ami}%
  \BibitemOpen
  \bibfield  {author} {\bibinfo {author} {\bibfnamefont {T.}~\bibnamefont
  {Akahori}}, \bibinfo {author} {\bibfnamefont {D.}~\bibnamefont {Ryu}}, \ and\
  \bibinfo {author} {\bibfnamefont {B.~M.}\ \bibnamefont {Gaensler}},\
  }\bibfield  {title} {\enquote {\bibinfo {title} {Fast radio bursts as probes
  of magnetic fields in the intergalactic medium},}\ }\href {\doibase
  https://doi.org/10.3847/0004-637X/824/2/105} {\bibfield  {journal} {\bibinfo
  {journal} {Astrophys. J.}\ }\textbf {\bibinfo {volume} {824}},\ \bibinfo
  {pages} {105} (\bibinfo {year} {2016})}\BibitemShut {NoStop}%
\bibitem [{\citenamefont {Deng}\ and\ \citenamefont {Zhang}(2014)}]{Deng_2014}%
  \BibitemOpen
  \bibfield  {author} {\bibinfo {author} {\bibfnamefont {W.}~\bibnamefont
  {Deng}}\ and\ \bibinfo {author} {\bibfnamefont {B.}~\bibnamefont {Zhang}},\
  }\bibfield  {title} {\enquote {\bibinfo {title} {Cosmological implications of
  fast radio burst/gamma-ray burst associations},}\ }\href {\doibase
  10.1088/2041-8205/783/2/L35} {\bibfield  {journal} {\bibinfo  {journal}
  {Astrophys. J. Lett.}\ }\textbf {\bibinfo {volume} {783}},\ \bibinfo {pages}
  {L35} (\bibinfo {year} {2014})}\BibitemShut {NoStop}%
\bibitem [{\citenamefont {Ravi}(2019)}]{Ravi:2018ose}%
  \BibitemOpen
  \bibfield  {author} {\bibinfo {author} {\bibfnamefont {V.}~\bibnamefont
  {Ravi}},\ }\bibfield  {title} {\enquote {\bibinfo {title} {{Measuring the
  Circumgalactic and Intergalactic Baryon Contents with Fast Radio Bursts}},}\
  }\href {\doibase 10.3847/1538-4357/aafb30} {\bibfield  {journal} {\bibinfo
  {journal} {Astrophys. J.}\ }\textbf {\bibinfo {volume} {872}},\ \bibinfo
  {pages} {88} (\bibinfo {year} {2019})}\BibitemShut {NoStop}%
\bibitem [{\citenamefont {Mu\~noz}\ and\ \citenamefont
  {Loeb}(2018)}]{Munoz:2018mll}%
  \BibitemOpen
  \bibfield  {author} {\bibinfo {author} {\bibfnamefont {J.~B.}\ \bibnamefont
  {Mu\~noz}}\ and\ \bibinfo {author} {\bibfnamefont {A.}~\bibnamefont {Loeb}},\
  }\bibfield  {title} {\enquote {\bibinfo {title} {Finding the missing baryons
  with fast radio bursts and sunyaev-zeldovich maps},}\ }\href {\doibase
  10.1103/PhysRevD.98.103518} {\bibfield  {journal} {\bibinfo  {journal} {Phys.
  Rev. D}\ }\textbf {\bibinfo {volume} {98}},\ \bibinfo {pages} {103518}
  (\bibinfo {year} {2018})}\BibitemShut {NoStop}%
\bibitem [{\citenamefont {Li}\ \emph {et~al.}(2019)\citenamefont {Li},
  \citenamefont {Gao}, \citenamefont {Wei}, \citenamefont {Yang}, \citenamefont
  {Zhang},\ and\ \citenamefont {Zhu}}]{li2019cosmology}%
  \BibitemOpen
  \bibfield  {author} {\bibinfo {author} {\bibfnamefont {Z.~X.}\ \bibnamefont
  {Li}}, \bibinfo {author} {\bibfnamefont {H.}~\bibnamefont {Gao}}, \bibinfo
  {author} {\bibfnamefont {J.~J.}\ \bibnamefont {Wei}}, \bibinfo {author}
  {\bibfnamefont {Y.~P.}\ \bibnamefont {Yang}}, \bibinfo {author}
  {\bibfnamefont {B.}~\bibnamefont {Zhang}}, \ and\ \bibinfo {author}
  {\bibfnamefont {Z.~H.}\ \bibnamefont {Zhu}},\ }\bibfield  {title} {\enquote
  {\bibinfo {title} {Cosmology-independent estimate of the fraction of baryon
  mass in the igm from fast radio burst observations},}\ }\href {\doibase
  10.3847/1538-4357/ab18fe} {\bibfield  {journal} {\bibinfo  {journal}
  {Astrophys. J.}\ }\textbf {\bibinfo {volume} {876}},\ \bibinfo {pages} {146}
  (\bibinfo {year} {2019})}\BibitemShut {NoStop}%
\bibitem [{\citenamefont {Li}\ \emph {et~al.}(2020)\citenamefont {Li},
  \citenamefont {Gao}, \citenamefont {Wei}, \citenamefont {Yang}, \citenamefont
  {Zhang},\ and\ \citenamefont {Zhu}}]{li2020cosmology}%
  \BibitemOpen
  \bibfield  {author} {\bibinfo {author} {\bibfnamefont {Z.~X.}\ \bibnamefont
  {Li}}, \bibinfo {author} {\bibfnamefont {H.}~\bibnamefont {Gao}}, \bibinfo
  {author} {\bibfnamefont {J.~J.}\ \bibnamefont {Wei}}, \bibinfo {author}
  {\bibfnamefont {Y.~P.}\ \bibnamefont {Yang}}, \bibinfo {author}
  {\bibfnamefont {B.}~\bibnamefont {Zhang}}, \ and\ \bibinfo {author}
  {\bibfnamefont {Z.~H.}\ \bibnamefont {Zhu}},\ }\bibfield  {title} {\enquote
  {\bibinfo {title} {Cosmology-insensitive estimate of igm baryon mass fraction
  from five localized fast radio bursts},}\ }\href {\doibase
  10.1093/mnrasl/slaa070} {\bibfield  {journal} {\bibinfo  {journal} {Mon. Not.
  Roy. Astron. Soc.}\ }\textbf {\bibinfo {volume} {496}},\ \bibinfo {pages}
  {L28--L32} (\bibinfo {year} {2020})}\BibitemShut {NoStop}%
\bibitem [{\citenamefont {Walters}\ \emph {et~al.}(2019)\citenamefont
  {Walters}, \citenamefont {Ma}, \citenamefont {Sievers},\ and\ \citenamefont
  {Weltman}}]{walters2019probing}%
  \BibitemOpen
  \bibfield  {author} {\bibinfo {author} {\bibfnamefont {A.}~\bibnamefont
  {Walters}}, \bibinfo {author} {\bibfnamefont {Y.~Z.}\ \bibnamefont {Ma}},
  \bibinfo {author} {\bibfnamefont {J.}~\bibnamefont {Sievers}}, \ and\
  \bibinfo {author} {\bibfnamefont {A.}~\bibnamefont {Weltman}},\ }\bibfield
  {title} {\enquote {\bibinfo {title} {Probing diffuse gas with fast radio
  bursts},}\ }\href {\doibase 10.1103/PhysRevD.100.103519} {\bibfield
  {journal} {\bibinfo  {journal} {Phys. Rev. D}\ }\textbf {\bibinfo {volume}
  {100}},\ \bibinfo {pages} {103519} (\bibinfo {year} {2019})}\BibitemShut
  {NoStop}%
\bibitem [{\citenamefont {Wei}\ \emph {et~al.}(2019)\citenamefont {Wei},
  \citenamefont {Li}, \citenamefont {Gao},\ and\ \citenamefont
  {Wu}}]{wei2019constraining}%
  \BibitemOpen
  \bibfield  {author} {\bibinfo {author} {\bibfnamefont {J.~J.}\ \bibnamefont
  {Wei}}, \bibinfo {author} {\bibfnamefont {Z.~X.}\ \bibnamefont {Li}},
  \bibinfo {author} {\bibfnamefont {H.}~\bibnamefont {Gao}}, \ and\ \bibinfo
  {author} {\bibfnamefont {X.~F.}\ \bibnamefont {Wu}},\ }\bibfield  {title}
  {\enquote {\bibinfo {title} {Constraining the evolution of the baryon
  fraction in the igm with frb and h (z) data},}\ }\href {\doibase
  10.1088/1475-7516/2019/09/039} {\bibfield  {journal} {\bibinfo  {journal}
  {JCAP}\ }\textbf {\bibinfo {volume} {09}},\ \bibinfo {pages} {039} (\bibinfo
  {year} {2019})}\BibitemShut {NoStop}%
\bibitem [{\citenamefont {Macquart}\ \emph {et~al.}(2020)\citenamefont
  {Macquart}, \citenamefont {Prochaska}, \citenamefont {McQuinn}, \citenamefont
  {Bannister}, \citenamefont {Bhandari}, \citenamefont {Day}, \citenamefont
  {Deller}, \citenamefont {Ekers}, \citenamefont {James}, \citenamefont
  {Marnoch} \emph {et~al.}}]{macquart2020census}%
  \BibitemOpen
  \bibfield  {author} {\bibinfo {author} {\bibfnamefont {J.~P.}\ \bibnamefont
  {Macquart}}, \bibinfo {author} {\bibfnamefont {J.~X.}\ \bibnamefont
  {Prochaska}}, \bibinfo {author} {\bibfnamefont {M.}~\bibnamefont {McQuinn}},
  \bibinfo {author} {\bibfnamefont {K.~W.}\ \bibnamefont {Bannister}}, \bibinfo
  {author} {\bibfnamefont {S.}~\bibnamefont {Bhandari}}, \bibinfo {author}
  {\bibfnamefont {C.~K.}\ \bibnamefont {Day}}, \bibinfo {author} {\bibfnamefont
  {A.~T.}\ \bibnamefont {Deller}}, \bibinfo {author} {\bibfnamefont {R.~D.}\
  \bibnamefont {Ekers}}, \bibinfo {author} {\bibfnamefont {C.~W.}\ \bibnamefont
  {James}}, \bibinfo {author} {\bibfnamefont {L.}~\bibnamefont {Marnoch}},
  \emph {et~al.},\ }\bibfield  {title} {\enquote {\bibinfo {title} {A census of
  baryons in the universe from localized fast radio bursts},}\ }\href {\doibase
  10.1038/s41586-020-2300-2} {\bibfield  {journal} {\bibinfo  {journal}
  {Nature}\ }\textbf {\bibinfo {volume} {581}},\ \bibinfo {pages} {391--395}
  (\bibinfo {year} {2020})}\BibitemShut {NoStop}%
\bibitem [{\citenamefont {Zhao}\ \emph {et~al.}(2022)\citenamefont {Zhao},
  \citenamefont {Zhang}, \citenamefont {Li}, \citenamefont {Zhang},\ and\
  \citenamefont {Zhang}}]{Zhao:2022yiv}%
  \BibitemOpen
  \bibfield  {author} {\bibinfo {author} {\bibfnamefont {Z.~W.}\ \bibnamefont
  {Zhao}}, \bibinfo {author} {\bibfnamefont {J.~G.}\ \bibnamefont {Zhang}},
  \bibinfo {author} {\bibfnamefont {Y.}~\bibnamefont {Li}}, \bibinfo {author}
  {\bibfnamefont {J.~F.}\ \bibnamefont {Zhang}}, \ and\ \bibinfo {author}
  {\bibfnamefont {X.}~\bibnamefont {Zhang}},\ }\bibfield  {title} {\enquote
  {\bibinfo {title} {{FRB dark sirens: Measuring the Hubble constant with
  unlocalized fast radio bursts}},}\ }\href@noop {} {\  (\bibinfo {year}
  {2022})},\ \Eprint {http://arxiv.org/abs/2212.13433} {arXiv:2212.13433
  [astro-ph.CO]} \BibitemShut {NoStop}%
\bibitem [{\citenamefont {Wei}\ and\ \citenamefont
  {Melia}(2023)}]{wei2023investigating}%
  \BibitemOpen
  \bibfield  {author} {\bibinfo {author} {\bibfnamefont {J.~J.}\ \bibnamefont
  {Wei}}\ and\ \bibinfo {author} {\bibfnamefont {F.}~\bibnamefont {Melia}},\
  }\bibfield  {title} {\enquote {\bibinfo {title} {Investigating cosmological
  models and the hubble tension using localized fast radio bursts},}\ }\href
  {\doibase 10.3847/1538-4357/acefb8} {\bibfield  {journal} {\bibinfo
  {journal} {Astrophys. J.}\ }\textbf {\bibinfo {volume} {955}},\ \bibinfo
  {pages} {101} (\bibinfo {year} {2023})}\BibitemShut {NoStop}%
\bibitem [{\citenamefont {Wu}\ \emph {et~al.}(2022)\citenamefont {Wu},
  \citenamefont {Zhang},\ and\ \citenamefont {Wang}}]{wu20228}%
  \BibitemOpen
  \bibfield  {author} {\bibinfo {author} {\bibfnamefont {Q.}~\bibnamefont
  {Wu}}, \bibinfo {author} {\bibfnamefont {G.~Q.}\ \bibnamefont {Zhang}}, \
  and\ \bibinfo {author} {\bibfnamefont {F.~Y.}\ \bibnamefont {Wang}},\
  }\bibfield  {title} {\enquote {\bibinfo {title} {An 8 per cent determination
  of the hubble constant from localized fast radio bursts},}\ }\href {\doibase
  10.1093/mnrasl/slac022} {\bibfield  {journal} {\bibinfo  {journal} {Mon. Not.
  Roy. Astron. Soc.}\ }\textbf {\bibinfo {volume} {515}},\ \bibinfo {pages}
  {L1--L5} (\bibinfo {year} {2022})}\BibitemShut {NoStop}%
\bibitem [{\citenamefont {Gao}\ \emph {et~al.}()\citenamefont {Gao},
  \citenamefont {Wu}, \citenamefont {Hu}, \citenamefont {Yi}, \citenamefont
  {Zhou},\ and\ \citenamefont {Wang}}]{gao2024measuring}%
  \BibitemOpen
  \bibfield  {author} {\bibinfo {author} {\bibfnamefont {D.~H.}\ \bibnamefont
  {Gao}}, \bibinfo {author} {\bibfnamefont {Q.}~\bibnamefont {Wu}}, \bibinfo
  {author} {\bibfnamefont {J.~P.}\ \bibnamefont {Hu}}, \bibinfo {author}
  {\bibfnamefont {S.~X.}\ \bibnamefont {Yi}}, \bibinfo {author} {\bibfnamefont
  {X.}~\bibnamefont {Zhou}}, \ and\ \bibinfo {author} {\bibfnamefont {F.~Y.}\
  \bibnamefont {Wang}},\ }\bibfield  {title} {\enquote {\bibinfo {title}
  {Measuring hubble constant using localized and unlocalized fast radio
  bursts},}\ }\href@noop {} {\ }\Eprint {http://arxiv.org/abs/2410.03994}
  {arXiv:2410.03994 [astro-ph.]} \BibitemShut {NoStop}%
\bibitem [{\citenamefont {W.}\ \emph {et~al.}(2025)\citenamefont {W.},
  \citenamefont {Gao},\ and\ \citenamefont {Fan}}]{wang2025probing}%
  \BibitemOpen
  \bibfield  {author} {\bibinfo {author} {\bibfnamefont {Y.~Y.}\ \bibnamefont
  {W.}}, \bibinfo {author} {\bibfnamefont {S.~J.}\ \bibnamefont {Gao}}, \ and\
  \bibinfo {author} {\bibfnamefont {Y.~Z.}\ \bibnamefont {Fan}},\ }\bibfield
  {title} {\enquote {\bibinfo {title} {Probing cosmology with 92 localized fast
  radio bursts and desi bao},}\ }\href {\doibase 10.3847/1538-4357/adade8}
  {\bibfield  {journal} {\bibinfo  {journal} {Astrophys. J.}\ }\textbf
  {\bibinfo {volume} {981}},\ \bibinfo {pages} {9} (\bibinfo {year}
  {2025})}\BibitemShut {NoStop}%
\bibitem [{\citenamefont {Kalita}\ \emph {et~al.}(2024)\citenamefont {Kalita},
  \citenamefont {Bhatporia},\ and\ \citenamefont {Weltman}}]{kalita2024fast}%
  \BibitemOpen
  \bibfield  {author} {\bibinfo {author} {\bibfnamefont {S.}~\bibnamefont
  {Kalita}}, \bibinfo {author} {\bibfnamefont {S.}~\bibnamefont {Bhatporia}}, \
  and\ \bibinfo {author} {\bibfnamefont {A.}~\bibnamefont {Weltman}},\
  }\bibfield  {title} {\enquote {\bibinfo {title} {Fast radio bursts as probes
  of the late-time universe: a new insight on the hubble tension},}\
  }\href@noop {} {\  (\bibinfo {year} {2024})},\ \Eprint
  {http://arxiv.org/abs/2410.01974} {arXiv:2410.01974 [astro-ph.CO]}
  \BibitemShut {NoStop}%
\bibitem [{\citenamefont {Gao}\ \emph {et~al.}(2024)\citenamefont {Gao},
  \citenamefont {Zhou}, \citenamefont {Du}, \citenamefont {Zou}, \citenamefont
  {Hu},\ and\ \citenamefont {Xu}}]{gao2024measurement}%
  \BibitemOpen
  \bibfield  {author} {\bibinfo {author} {\bibfnamefont {J.}~\bibnamefont
  {Gao}}, \bibinfo {author} {\bibfnamefont {Z.}~\bibnamefont {Zhou}}, \bibinfo
  {author} {\bibfnamefont {M.}~\bibnamefont {Du}}, \bibinfo {author}
  {\bibfnamefont {R.}~\bibnamefont {Zou}}, \bibinfo {author} {\bibfnamefont
  {J.}~\bibnamefont {Hu}}, \ and\ \bibinfo {author} {\bibfnamefont
  {L.}~\bibnamefont {Xu}},\ }\bibfield  {title} {\enquote {\bibinfo {title} {A
  measurement of hubble constant using cosmographic approach combining fast
  radio bursts and supernovae},}\ }\href {\doibase 10.1093/mnras/stad3708}
  {\bibfield  {journal} {\bibinfo  {journal} {Mon. Not. Roy. Astron. Soc.}\
  }\textbf {\bibinfo {volume} {527}},\ \bibinfo {pages} {7861--7870} (\bibinfo
  {year} {2024})}\BibitemShut {NoStop}%
\bibitem [{\citenamefont {Yang}\ \emph {et~al.}(2022)\citenamefont {Yang},
  \citenamefont {Wu},\ and\ \citenamefont {Wang}}]{yang2022finding}%
  \BibitemOpen
  \bibfield  {author} {\bibinfo {author} {\bibfnamefont {K.~B.}\ \bibnamefont
  {Yang}}, \bibinfo {author} {\bibfnamefont {Q.}~\bibnamefont {Wu}}, \ and\
  \bibinfo {author} {\bibfnamefont {F.~Y.}\ \bibnamefont {Wang}},\ }\bibfield
  {title} {\enquote {\bibinfo {title} {Finding the missing baryons in the
  intergalactic medium with localized fast radio bursts},}\ }\href {\doibase
  10.3847/2041-8213/aca145} {\bibfield  {journal} {\bibinfo  {journal}
  {Astrophys. J. Lett.}\ }\textbf {\bibinfo {volume} {940}},\ \bibinfo {pages}
  {L29} (\bibinfo {year} {2022})}\BibitemShut {NoStop}%
\bibitem [{\citenamefont {Lin}\ and\ \citenamefont
  {Zou}(2023)}]{lin2023probing}%
  \BibitemOpen
  \bibfield  {author} {\bibinfo {author} {\bibfnamefont {H.~N.}\ \bibnamefont
  {Lin}}\ and\ \bibinfo {author} {\bibfnamefont {R.}~\bibnamefont {Zou}},\
  }\bibfield  {title} {\enquote {\bibinfo {title} {Probing the baryon mass
  fraction in igm and its redshift evolution with fast radio bursts using
  bayesian inference method},}\ }\href {\doibase 10.1093/mnras/stad509}
  {\bibfield  {journal} {\bibinfo  {journal} {Mon. Not. Roy. Astron. Soc.}\
  }\textbf {\bibinfo {volume} {520}},\ \bibinfo {pages} {6237--6244} (\bibinfo
  {year} {2023})}\BibitemShut {NoStop}%
\bibitem [{\citenamefont {Wang}\ and\ \citenamefont {Wei}(2023)}]{wang20238}%
  \BibitemOpen
  \bibfield  {author} {\bibinfo {author} {\bibfnamefont {B.}~\bibnamefont
  {Wang}}\ and\ \bibinfo {author} {\bibfnamefont {J.~J.}\ \bibnamefont {Wei}},\
  }\bibfield  {title} {\enquote {\bibinfo {title} {An 8.0\% determination of
  the baryon fraction in the intergalactic medium from localized fast radio
  bursts},}\ }\href {\doibase 10.3847/1538-4357/acb2c8} {\bibfield  {journal}
  {\bibinfo  {journal} {Astrophys. J.}\ }\textbf {\bibinfo {volume} {944}},\
  \bibinfo {pages} {50} (\bibinfo {year} {2023})}\BibitemShut {NoStop}%
\bibitem [{\citenamefont {Fortunato}\ \emph {et~al.}(2023)\citenamefont
  {Fortunato}, \citenamefont {Hip\'olito-Ricaldi},\ and\ \citenamefont {dos
  Santos}}]{Fortunato:2023deh}%
  \BibitemOpen
  \bibfield  {author} {\bibinfo {author} {\bibfnamefont {J.~A.~S.}\
  \bibnamefont {Fortunato}}, \bibinfo {author} {\bibfnamefont {W.~S.}\
  \bibnamefont {Hip\'olito-Ricaldi}}, \ and\ \bibinfo {author} {\bibfnamefont
  {M.~V.}\ \bibnamefont {dos Santos}},\ }\bibfield  {title} {\enquote {\bibinfo
  {title} {{Cosmography from well-localized fast radio bursts}},}\ }\href
  {\doibase 10.1093/mnras/stad2856} {\bibfield  {journal} {\bibinfo  {journal}
  {Mon. Not. Roy. Astron. Soc.}\ }\textbf {\bibinfo {volume} {526}},\ \bibinfo
  {pages} {1773--1782} (\bibinfo {year} {2023})}\BibitemShut {NoStop}%
\bibitem [{\citenamefont {Lin}\ \emph {et~al.}(2023)\citenamefont {Lin},
  \citenamefont {Tang},\ and\ \citenamefont {Zou}}]{lin2023revised}%
  \BibitemOpen
  \bibfield  {author} {\bibinfo {author} {\bibfnamefont {H.~N.}\ \bibnamefont
  {Lin}}, \bibinfo {author} {\bibfnamefont {L.}~\bibnamefont {Tang}}, \ and\
  \bibinfo {author} {\bibfnamefont {R.}~\bibnamefont {Zou}},\ }\bibfield
  {title} {\enquote {\bibinfo {title} {Revised constraints on the photon mass
  from well-localized fast radio bursts},}\ }\href {\doibase
  10.1093/mnras/stad228} {\bibfield  {journal} {\bibinfo  {journal} {Mon. Not.
  Roy. Astron. Soc.}\ }\textbf {\bibinfo {volume} {520}},\ \bibinfo {pages}
  {1324--1331} (\bibinfo {year} {2023})}\BibitemShut {NoStop}%
\bibitem [{\citenamefont {Wang}\ \emph {et~al.}(2023)\citenamefont {Wang},
  \citenamefont {Wei}, \citenamefont {Wu},\ and\ \citenamefont
  {L\'opez-Corredoira}}]{Wang:2023fnn}%
  \BibitemOpen
  \bibfield  {author} {\bibinfo {author} {\bibfnamefont {B.}~\bibnamefont
  {Wang}}, \bibinfo {author} {\bibfnamefont {J.~J.}\ \bibnamefont {Wei}},
  \bibinfo {author} {\bibfnamefont {X.~F.}\ \bibnamefont {Wu}}, \ and\ \bibinfo
  {author} {\bibfnamefont {M.}~\bibnamefont {L\'opez-Corredoira}},\ }\bibfield
  {title} {\enquote {\bibinfo {title} {{Revisiting constraints on the photon
  rest mass with cosmological fast radio bursts}},}\ }\href {\doibase
  10.1088/1475-7516/2023/09/025} {\bibfield  {journal} {\bibinfo  {journal}
  {JCAP}\ }\textbf {\bibinfo {volume} {09}},\ \bibinfo {pages} {025} (\bibinfo
  {year} {2023})}\BibitemShut {NoStop}%
\bibitem [{\citenamefont {Kalita}(2024)}]{kalita2024constraining}%
  \BibitemOpen
  \bibfield  {author} {\bibinfo {author} {\bibfnamefont {S.}~\bibnamefont
  {Kalita}},\ }\bibfield  {title} {\enquote {\bibinfo {title} {Constraining
  fundamental constants with fast radio bursts: unveiling the role of energy
  scale},}\ }\href {\doibase 10.1093/mnrasl/slae062} {\bibfield  {journal}
  {\bibinfo  {journal} {Mon. Not. Roy. Astron. Soc.}\ }\textbf {\bibinfo
  {volume} {533}},\ \bibinfo {pages} {L57--L63} (\bibinfo {year}
  {2024})}\BibitemShut {NoStop}%
\bibitem [{\citenamefont {McQuinn}(2013)}]{mcquinn2013locating}%
  \BibitemOpen
  \bibfield  {author} {\bibinfo {author} {\bibfnamefont {M.}~\bibnamefont
  {McQuinn}},\ }\bibfield  {title} {\enquote {\bibinfo {title} {Locating the
  `missing' baryons with extragalactic dispersion measure estimates},}\ }\href
  {\doibase 10.1088/2041-8205/780/2/L33} {\bibfield  {journal} {\bibinfo
  {journal} {Astrophys. J. Lett.}\ }\textbf {\bibinfo {volume} {780}},\
  \bibinfo {pages} {L33} (\bibinfo {year} {2013})}\BibitemShut {NoStop}%
\bibitem [{\citenamefont {Zhang}\ \emph {et~al.}(2021)\citenamefont {Zhang},
  \citenamefont {Yan}, \citenamefont {Li}, \citenamefont {Zhang},\ and\
  \citenamefont {Wang}}]{zhang2021intergalactic}%
  \BibitemOpen
  \bibfield  {author} {\bibinfo {author} {\bibfnamefont {Z.~J.}\ \bibnamefont
  {Zhang}}, \bibinfo {author} {\bibfnamefont {K.}~\bibnamefont {Yan}}, \bibinfo
  {author} {\bibfnamefont {C.~M.}\ \bibnamefont {Li}}, \bibinfo {author}
  {\bibfnamefont {G.~Q.}\ \bibnamefont {Zhang}}, \ and\ \bibinfo {author}
  {\bibfnamefont {F.~Y.}\ \bibnamefont {Wang}},\ }\bibfield  {title} {\enquote
  {\bibinfo {title} {Intergalactic medium dispersion measures of fast radio
  bursts estimated from illustristng simulation and their cosmological
  applications},}\ }\href {\doibase 10.3847/1538-4357/abceb9} {\bibfield
  {journal} {\bibinfo  {journal} {Astrophys. J.}\ }\textbf {\bibinfo {volume}
  {906}},\ \bibinfo {pages} {49} (\bibinfo {year} {2021})}\BibitemShut
  {NoStop}%
\bibitem [{\citenamefont {Rohatgi}\ and\ \citenamefont
  {Ehsanes~Saleh}(2000)}]{Rohatgi}%
  \BibitemOpen
  \bibfield  {author} {\bibinfo {author} {\bibfnamefont {V.~K.}\ \bibnamefont
  {Rohatgi}}\ and\ \bibinfo {author} {\bibfnamefont {A.~K.~MD.}\ \bibnamefont
  {Ehsanes~Saleh}},\ }\href {\doibase https://doi.org/10.1002/9781118165676}
  {\emph {\bibinfo {title} {An Introduction to Probability and Statistics}}}\
  (\bibinfo  {publisher} {Wiley},\ \bibinfo {year} {2000})\BibitemShut
  {NoStop}%
\bibitem [{\citenamefont {Zhuge}\ \emph {et~al.}(2025)\citenamefont {Zhuge},
  \citenamefont {Kalomenopoulos},\ and\ \citenamefont {Zhang}}]{Zhuge:2025urk}%
  \BibitemOpen
  \bibfield  {author} {\bibinfo {author} {\bibfnamefont {Jiaming}\ \bibnamefont
  {Zhuge}}, \bibinfo {author} {\bibfnamefont {Marios}\ \bibnamefont
  {Kalomenopoulos}}, \ and\ \bibinfo {author} {\bibfnamefont {Bing}\
  \bibnamefont {Zhang}},\ }\bibfield  {title} {\enquote {\bibinfo {title}
  {{Hubble constant constraint using 117 FRBs with a more accurate probability
  density function for ${\rm DM}_{\rm diff}$}},}\ }\href@noop {} {\  (\bibinfo
  {year} {2025})},\ \Eprint {http://arxiv.org/abs/2508.05161} {arXiv:2508.05161
  [astro-ph.CO]} \BibitemShut {NoStop}%
\bibitem [{\citenamefont {Qiang}\ and\ \citenamefont
  {Wei}(2021)}]{qiang2021effect}%
  \BibitemOpen
  \bibfield  {author} {\bibinfo {author} {\bibfnamefont {D.~C.}\ \bibnamefont
  {Qiang}}\ and\ \bibinfo {author} {\bibfnamefont {H.}~\bibnamefont {Wei}},\
  }\bibfield  {title} {\enquote {\bibinfo {title} {Effect of redshift
  distributions of fast radio bursts on cosmological constraints},}\ }\href
  {\doibase 10.1103/PhysRevD.103.083536} {\bibfield  {journal} {\bibinfo
  {journal} {Phys. Rev. D}\ }\textbf {\bibinfo {volume} {103}},\ \bibinfo
  {pages} {083536} (\bibinfo {year} {2021})}\BibitemShut {NoStop}%
\bibitem [{\citenamefont {Marcote}\ \emph {et~al.}(2020)\citenamefont {Marcote}
  \emph {et~al.}}]{Marcote:2020ljw}%
  \BibitemOpen
  \bibfield  {author} {\bibinfo {author} {\bibfnamefont {B.}~\bibnamefont
  {Marcote}} \emph {et~al.},\ }\bibfield  {title} {\enquote {\bibinfo {title}
  {{A repeating fast radio burst source localized to a nearby spiral
  galaxy}},}\ }\href {\doibase 10.1038/s41586-019-1866-z} {\bibfield  {journal}
  {\bibinfo  {journal} {Nature}\ }\textbf {\bibinfo {volume} {577}},\ \bibinfo
  {pages} {190--194} (\bibinfo {year} {2020})}\BibitemShut {NoStop}%
\bibitem [{\citenamefont {Tendulkar}\ \emph {et~al.}(2017)\citenamefont
  {Tendulkar} \emph {et~al.}}]{Tendulkar:2017vuq}%
  \BibitemOpen
  \bibfield  {author} {\bibinfo {author} {\bibfnamefont {S.~P.}\ \bibnamefont
  {Tendulkar}} \emph {et~al.},\ }\bibfield  {title} {\enquote {\bibinfo {title}
  {{The Host Galaxy and Redshift of the Repeating Fast Radio Burst FRB
  121102}},}\ }\href {\doibase 10.3847/2041-8213/834/2/L7} {\bibfield
  {journal} {\bibinfo  {journal} {Astrophys. J. Lett.}\ }\textbf {\bibinfo
  {volume} {834}},\ \bibinfo {pages} {L7} (\bibinfo {year} {2017})}\BibitemShut
  {NoStop}%
\bibitem [{\citenamefont {Bannister}\ \emph {et~al.}(2019)\citenamefont
  {Bannister} \emph {et~al.}}]{Bannister:2019iju}%
  \BibitemOpen
  \bibfield  {author} {\bibinfo {author} {\bibfnamefont {K.~W.}\ \bibnamefont
  {Bannister}} \emph {et~al.},\ }\bibfield  {title} {\enquote {\bibinfo {title}
  {{A single fast radio burst localized to a massive galaxy at cosmological
  distance}},}\ }\href {\doibase 10.1126/science.aaw5903} {\bibfield  {journal}
  {\bibinfo  {journal} {Science}\ }\textbf {\bibinfo {volume} {365}},\ \bibinfo
  {pages} {6453} (\bibinfo {year} {2019})}\BibitemShut {NoStop}%
\bibitem [{\citenamefont {Zhang}\ \emph {et~al.}(2020)\citenamefont {Zhang},
  \citenamefont {Yu}, \citenamefont {He},\ and\ \citenamefont
  {Wang}}]{zhang2020dispersion}%
  \BibitemOpen
  \bibfield  {author} {\bibinfo {author} {\bibfnamefont {G.~Q.}\ \bibnamefont
  {Zhang}}, \bibinfo {author} {\bibfnamefont {H.}~\bibnamefont {Yu}}, \bibinfo
  {author} {\bibfnamefont {J.~H.}\ \bibnamefont {He}}, \ and\ \bibinfo {author}
  {\bibfnamefont {F.~Y.}\ \bibnamefont {Wang}},\ }\bibfield  {title} {\enquote
  {\bibinfo {title} {Dispersion measures of fast radio burst host galaxies
  derived from illustristng simulation},}\ }\href {\doibase
  10.3847/1538-4357/abaa4a} {\bibfield  {journal} {\bibinfo  {journal}
  {Astrophys. J.}\ }\textbf {\bibinfo {volume} {900}},\ \bibinfo {pages} {170}
  (\bibinfo {year} {2020})}\BibitemShut {NoStop}%
\bibitem [{\citenamefont {Ioka}(2003)}]{ioka2003cosmic}%
  \BibitemOpen
  \bibfield  {author} {\bibinfo {author} {\bibfnamefont {K.}~\bibnamefont
  {Ioka}},\ }\bibfield  {title} {\enquote {\bibinfo {title} {The cosmic
  dispersion measure from gamma-ray burst afterglows: probing the reionization
  history and the burst environment},}\ }\href {\doibase 10.1086/380598}
  {\bibfield  {journal} {\bibinfo  {journal} {Astrophys. J.}\ }\textbf
  {\bibinfo {volume} {598}},\ \bibinfo {pages} {L79} (\bibinfo {year}
  {2003})}\BibitemShut {NoStop}%
\bibitem [{\citenamefont {Inoue}(2004)}]{inoue2004probing}%
  \BibitemOpen
  \bibfield  {author} {\bibinfo {author} {\bibfnamefont {S.}~\bibnamefont
  {Inoue}},\ }\bibfield  {title} {\enquote {\bibinfo {title} {Probing the
  cosmic reionization history and local environment of gamma-ray bursts through
  radio dispersion},}\ }\href {\doibase 10.1111/j.1365-2966.2004.07359.x}
  {\bibfield  {journal} {\bibinfo  {journal} {Mon. Not. Roy. Astron. Soc.}\
  }\textbf {\bibinfo {volume} {348}},\ \bibinfo {pages} {999--1008} (\bibinfo
  {year} {2004})}\BibitemShut {NoStop}%
\bibitem [{\citenamefont {Aghanim}\ \emph {et~al.}(2020)\citenamefont
  {Aghanim}, \citenamefont {Akrami}, \citenamefont {Ashdown}, \citenamefont
  {Aumont}, \citenamefont {Baccigalupi}, \citenamefont {Ballardini},
  \citenamefont {Banday}, \citenamefont {Barreiro}, \citenamefont {Bartolo},
  \citenamefont {Basak} \emph {et~al.}}]{aghanim2020planck}%
  \BibitemOpen
  \bibfield  {author} {\bibinfo {author} {\bibfnamefont {N.}~\bibnamefont
  {Aghanim}}, \bibinfo {author} {\bibfnamefont {Y.}~\bibnamefont {Akrami}},
  \bibinfo {author} {\bibfnamefont {M.}~\bibnamefont {Ashdown}}, \bibinfo
  {author} {\bibfnamefont {J.}~\bibnamefont {Aumont}}, \bibinfo {author}
  {\bibfnamefont {C.}~\bibnamefont {Baccigalupi}}, \bibinfo {author}
  {\bibfnamefont {M.}~\bibnamefont {Ballardini}}, \bibinfo {author}
  {\bibfnamefont {A.~J.}\ \bibnamefont {Banday}}, \bibinfo {author}
  {\bibfnamefont {R.~B.}\ \bibnamefont {Barreiro}}, \bibinfo {author}
  {\bibfnamefont {N.}~\bibnamefont {Bartolo}}, \bibinfo {author} {\bibfnamefont
  {S.}~\bibnamefont {Basak}},  \emph {et~al.},\ }\bibfield  {title} {\enquote
  {\bibinfo {title} {Planck 2018 results-vi. cosmological parameters},}\ }\href
  {\doibase 10.1051/0004-6361/201833910} {\bibfield  {journal} {\bibinfo
  {journal} {Astron. \& Astrophys.}\ }\textbf {\bibinfo {volume} {641}},\
  \bibinfo {pages} {A6} (\bibinfo {year} {2020})}\BibitemShut {NoStop}%
\bibitem [{\citenamefont {Fukugita}\ \emph {et~al.}(1998)\citenamefont
  {Fukugita}, \citenamefont {Hogan},\ and\ \citenamefont
  {Peebles}}]{Fukugita:1997bi}%
  \BibitemOpen
  \bibfield  {author} {\bibinfo {author} {\bibfnamefont {M.}~\bibnamefont
  {Fukugita}}, \bibinfo {author} {\bibfnamefont {C.~J.}\ \bibnamefont {Hogan}},
  \ and\ \bibinfo {author} {\bibfnamefont {P.~J.~E.}\ \bibnamefont {Peebles}},\
  }\bibfield  {title} {\enquote {\bibinfo {title} {{The Cosmic baryon
  budget}},}\ }\href {\doibase 10.1086/306025} {\bibfield  {journal} {\bibinfo
  {journal} {Astrophys. J.}\ }\textbf {\bibinfo {volume} {503}},\ \bibinfo
  {pages} {518} (\bibinfo {year} {1998})}\BibitemShut {NoStop}%
\bibitem [{\citenamefont {Foreman-Mackey}\ \emph {et~al.}(2013)\citenamefont
  {Foreman-Mackey}, \citenamefont {Hogg}, \citenamefont {Lang},\ and\
  \citenamefont {Goodman}}]{foreman2013emcee}%
  \BibitemOpen
  \bibfield  {author} {\bibinfo {author} {\bibfnamefont {D.}~\bibnamefont
  {Foreman-Mackey}}, \bibinfo {author} {\bibfnamefont {D.~W.}\ \bibnamefont
  {Hogg}}, \bibinfo {author} {\bibfnamefont {D.}~\bibnamefont {Lang}}, \ and\
  \bibinfo {author} {\bibfnamefont {J.}~\bibnamefont {Goodman}},\ }\bibfield
  {title} {\enquote {\bibinfo {title} {emcee: the mcmc hammer},}\ }\href
  {\doibase 10.1086/670067} {\bibfield  {journal} {\bibinfo  {journal}
  {Publications of the Astronomical Society of the Pacific}\ }\textbf {\bibinfo
  {volume} {125}},\ \bibinfo {pages} {306} (\bibinfo {year}
  {2013})}\BibitemShut {NoStop}%
\bibitem [{\citenamefont {Yao}\ \emph {et~al.}(2017)\citenamefont {Yao},
  \citenamefont {Manchester},\ and\ \citenamefont {Wang}}]{yao2017new}%
  \BibitemOpen
  \bibfield  {author} {\bibinfo {author} {\bibfnamefont {J.~M.}\ \bibnamefont
  {Yao}}, \bibinfo {author} {\bibfnamefont {R.~N.}\ \bibnamefont {Manchester}},
  \ and\ \bibinfo {author} {\bibfnamefont {N.}~\bibnamefont {Wang}},\
  }\bibfield  {title} {\enquote {\bibinfo {title} {A new electron-density model
  for estimation of pulsar and frb distances},}\ }\href {\doibase
  10.3847/1538-4357/835/1/29} {\bibfield  {journal} {\bibinfo  {journal}
  {Astrophys. J.}\ }\textbf {\bibinfo {volume} {835}},\ \bibinfo {pages} {29}
  (\bibinfo {year} {2017})}\BibitemShut {NoStop}%
\bibitem [{\citenamefont {Prochaska}\ and\ \citenamefont
  {Zheng}(2019)}]{prochaska2019probing}%
  \BibitemOpen
  \bibfield  {author} {\bibinfo {author} {\bibfnamefont {J.~X.}\ \bibnamefont
  {Prochaska}}\ and\ \bibinfo {author} {\bibfnamefont {Y.}~\bibnamefont
  {Zheng}},\ }\bibfield  {title} {\enquote {\bibinfo {title} {Probing galactic
  haloes with fast radio bursts},}\ }\href {\doibase 10.1093/mnras/stz261}
  {\bibfield  {journal} {\bibinfo  {journal} {Mon. Not. Roy. Astron. Soc.}\
  }\textbf {\bibinfo {volume} {485}},\ \bibinfo {pages} {648--665} (\bibinfo
  {year} {2019})}\BibitemShut {NoStop}%
\end{thebibliography}%
\end{document}